\documentclass[floatfix,showpacs,preprintnumbers,amsmath,amssymb]{revtex4}

\usepackage{graphicx}
\usepackage{dcolumn}
\usepackage{bm}
\begin{document}
\title{Charged current anti-neutrino reactions from $^{12}$C at MiniBooNE energies}
\author{M. Sajjad Athar, Shakeb Ahmad and S. K. Singh}
\email{pht13sks@rediffmail.com}
\affiliation{Department of Physics, Aligarh Muslim University, Aligarh-202 002, India.}
\date{\today}
\begin{abstract}
A study of charged current induced anti-neutrino interactions from nuclei has been done for the intermediate energy ($<2$GeV) anti-neutrinos and applied to $^{12}$C, relevant for ongoing experiment by MiniBooNE collaboration. The calculations have been done for the quasielastic and inelastic lepton production as well as for the incoherent and the coherent pion production processes. The calculations are done in local density approximation. In the case of the quasielastic reaction the effects of Pauli blocking, Fermi motion effects, renormalization of weak transition strengths in nuclear medium and the Coulomb distortion of the outgoing lepton have been taken into account. For the inelastic processes the calculations have been done in the $\Delta$ dominance model and take into account the effect of Pauli blocking, Fermi motion of the nucleon and renormalization of $\Delta$ properties in a nuclear medium. The effect of final state interactions of pions is also taken into account. The numerical results for the total cross sections for the charged current quasielastic scattering and incoherent pion production processes are compared with earlier experimental results available in Freon and Freon-Propane. It is found that nuclear medium effects give strong reduction in the cross sections leading to satisfactory agreement with the available data.
\end{abstract}
\maketitle 
\section{Introduction}
Recent interest in the study of CP violation in leptonic sector makes it desirable that anti-neutrino reactions be studied in the same kinematic regions as neutrino reactions~\cite{cp}. While there are many experimental studies made for neutrino reactions, there are very few studies made for anti-neutrino reactions from nucleons and nuclei. In the low energy region, the experimental studies with anti-neutrinos were first made using anti-neutrinos from nuclear reactors at Savannah river facility~\cite{reins} but recently similar experiments have been done by CHOOZ~\cite{appl} and KamLAND~\cite{kam} collaborations. In the intermediate energy region of few GeV, the earlier experiments performed with anti-neutrino beams at CERN and Serpukhov laboratories, provided the cross sections for various quasielastic~\cite{bon}-\cite{bru} and inelastic~\cite{bolognese} reactions. In the new generation of anti-neutrino experiments, the experiment performed at LSND~\cite{aur} with anti-neutrinos in the energy region of few hundreds MeV has attracted much attention due to its role in the study of neutrino oscillations. There are now many experiments being done to study the neutrino and anti-neutrino reaction cross section from various nuclei in the intermediate energy region of neutrinos and anti-neutrinos. One of the present experiments being done at Fermi Lab by the MiniBooNE collaboration is designed to study specifically the nuclear reactions induced by anti-neutrinos~\cite{a1}-\cite{boone1}.

At the MiniBooNE energies, the contribution to the anti-neutrino nucleus cross section comes mainly from the charged current quasielastic, neutral current elastic, and charged and neutral current one pion production processes. In the intermediate energy region, there are only few experimental measurements on ${\bar\nu}_\mu$ quasielastic scattering cross sections which are made using bubble chambers at CERN and Serpukhov~\cite{bon}-\cite{bru}. They have limited statistics. Moreover, there is lack of experimental data particularly for $E_{{\bar\nu}_\mu}<$1GeV~\cite{bon}-\cite{bru}. These experiments on quasielastic processes have been generally analysed in a Fermi gas model using a global Fermi momentum and constant binding energy as parameters~\cite{smith}-\cite{gaisser}. There are, however, various other theoretical calculations for the quasielastic processes which make use of various nuclear models like Shell Model with pairing correlations, Random Phase Approximation, Relativistic Mean Field Approximation, etc.~\cite{kuramoto}-\cite{udias}. Recently many calculations have been done for these processes in order to better understand the nuclear model dependencies of these cross sections~\cite{a2}-\cite{mosel}. However, these calculations are applied mainly to study the neutrino reactions and not much attention has been given to the anti-neutrino reactions. 

In the case of inelastic anti-neutrino nuclear reactions pion production processes have been studied. These are generally studied in a $\Delta$ dominance model in which pions are dominantly produced through the excitation of $\Delta$ and its subsequent decays leading to pions~\cite{mosel}-\cite{coh}. In case of nuclear production, there are coherent and incoherent processes through which pions are produced. The coherent reactions are more forward peaked than the incoherent reactions and therefore provide a better place to study the nuclear dynamics by minimizing the $q^2$ dependence of various transition from factors on the nuclear cross sections. The ratio of coherent to incoherent production of pions in anti-neutrino reaction cross sections is larger than the inelastic neutrino reactions. The major difference between these reactions is the production of hyperons which is allowed in the qusielastic reactions with anti-neutrino but are forbidden in the case of neutrino reactions~\cite{hyp}. These strangeness changing quasielastic reactions can also lead to pion productions through the decay of hyperons which can be important in the energy region of MiniBooNE~\cite{zeller} and K2K~\cite{k2k} experiments. 

The nuclear effects in the quasielastic as well as in the inelastic anti-neutrino reactions have not been analyzed as extensively in the literature as in the case of neutrino reactions. Such an analysis is important for making CP violation studies in the neutrino sector relevant for the next generation neutrino oscillation experiments.

In this paper, we study the charged current quasielastic and inelastic reactions induced by anti-neutrinos from nuclei. In the case of inelastic reactions, we study the inelastic lepton production as well as the coherent and the incoherent production of pions. The matrix elements for the quasielastic and the inelastic reactions from free nucleons are written in the standard model using the transition form factors which are determined from the analysis of experimental data available from the electron and neutrino scattering experiments from nucleon and deuteron targets in the relevant energy region~\cite{a2}. Nuclear effects, like Pauli blocking, Fermi motion etc.of initial nucleon(and also final nucleon in the case of quasielastic reactions) are taken into account in the local density approximation~\cite{singh},~\cite{ruso}. The coherent and the incoherent production of leptons accompanied by a pion is calculated in a $\Delta$ dominance model, which has been earlier applied to study the neutrino induced charged current 1$\pi^+$ production process~\cite{prd1} and the charged and neutral current neutrino induced coherent pion production processes~\cite{a6},~\cite{coh}. The renormalization of the $\Delta$ properties like modifications in mass and width in nuclear medium have been taken into account~\cite{a4}. In case of incoherent reactions, the effect of final state interaction of pions with the residual nucleus has been considered in eikonal approximation using probabilities per unit length for pion absorption given by Vicente Vacas et al.~\cite{a5} as the basic input. In the case of coherent reactions, the final state interaction of pions has been treated by calculating the distortion of pion in the optical potential using an eikonal approximation~\cite{a6}.

In section-II, the formalism for the charged current quasielastic lepton production in the local Fermi gas model including RPA correlations has been discussed. In section-III, we present the formalism for charged current inelastic reactions from the nucleons in the $\Delta$ dominance model and describe the nuclear medium and the final state interaction effects. In this section, we have also discussed the charged current incoherent and coherent lepton production accompanied by a pion, charged current incoherent 1$\pi^-$ production, and quasielastic like lepton production. In section-IV, we present and discuss the numerical results. In section-V, we provide a summary and conclusion of the present work. 
\section{Charged current quasielastic reactions}
In this section we derive an expression for the total scattering cross section $\sigma(E_{\bar\nu})$ for the charged current(CC) reaction 
\begin{equation}
{\bar{\nu}}_\mu +  ~_{Z_i}^{A}X  \rightarrow \mu^+ +  ~_{Z_f}^{A}Y,
\end{equation}
where $Z_i(Z_f)$ is the charge of initial(final) nucleus. 

The basic ${\bar{\nu}}_\mu$-nucleon quasielastic reaction taking place in $_{Z_i}^{A}X$ nucleus is on the proton target i.e.
\begin{equation}
{\bar{\nu}}_\mu(k) + p(p) \rightarrow \mu^+(k^\prime) + n(p^\prime),
\end{equation}
where $k$ and $p$ are the four momenta of the incoming anti-neutrino and proton and $k^\prime$ and $p^\prime$ are the four momenta of the outgoing muon and neutron respectively.

The matrix element for the basic anti-neutrino charged current process on free nucleon (Eq.2) is written as
\begin{equation}
T = \frac{G_F}{\sqrt{2}}\cos{\theta_c} ~l_{\mu} ~J^{\mu},
\end{equation}
where
\begin{eqnarray}
l_{\mu}&=&\bar{v}(k)\gamma_\mu(1-\gamma_5){v}(k^\prime)\nonumber\\ 
J^\mu&=&\bar{u}(p^\prime)[F_{1}^V(q^2)\gamma^\mu+F_{2}^V(q^2)i{\sigma^{\mu\nu}}{\frac{q_\nu}{2M}}- F_{A}^V(q^2)\gamma^\mu\gamma_5 - F_{P}^V(q^2)q^\mu\gamma_5 ]u(p).
\end{eqnarray}
$q(=k-k^\prime)$ is the four momentum transfer and the expressions for $F_1^V(q^2)$, $F_2^V(q^2)$, $F_P^V(q^2)$ and $F_A^V(q^2)$ are the weak nuclear form factors taken from Bradford et al.~\cite{budd}(BBBA05).

The double differential cross section $d^2\sigma_{\text{free}}(E_\mu,|\vec k^\prime|)$ for the basic reaction described in Eq.(2) is then written as
\begin{equation}
d^2\sigma_{\text{free}}(E_\mu,|\vec k^\prime|)=\frac{{|\vec k^\prime|}^2}{4\pi E_{{\bar{\nu}}_\mu} E_\mu}\frac{M_n M_p}{E_n E_p}{\bar\Sigma}\Sigma|T|^2\delta[q_0+E_p-E_n]
\end{equation}
where ${\bar{\Sigma}}{\Sigma} {|T|^2}$ is the square modulus of the transition amplitude given by
\begin{eqnarray}
{\bar{\Sigma}}{\Sigma} {|T|^2}&=&\frac{{G_F}^2\cos^2{\theta_c}}{2}L_{\mu\nu}J^{\mu\nu}~~\mbox{with}\nonumber\\
L_{\mu\nu}&=&8\left[k_\mu {k^\prime}_\nu + k_\nu {k^\prime}_\mu - k.k^\prime g_{\mu\nu} - i\epsilon_{\mu\nu\lambda\sigma}{k}^\lambda{k^\prime}^\sigma\right]\nonumber\\
J^{\mu\nu}&=&\sum{J^\mu}^\dagger J^\nu
\end{eqnarray}
$q_0(q_0=E_{{\bar{\nu}}_\mu} - E_\mu)$ is the energy transferred to the nucleon.

In a nucleus, the anti-neutrino scatters from a proton moving in the finite nucleus of proton density $\rho_p(r)$, with a local occupation number $n_p({\vec{p}},{\vec{r}})$. In the local density approximation the differential scattering cross section for the nucleon is written as
\begin{equation}
d^2\sigma_{A}(E_\mu,|\vec k^\prime|)=\int 2d{\vec r}d{\vec p}\frac{1}{(2\pi)^3}n_p({\vec p},{\vec r})d^2\sigma_{\text{free}}(E_\mu,|\vec k^\prime|)
\end{equation}
The proton energy $E_p$ and neutron energy $E_n$ are replaced by $E_p(|\vec p|)$ and $E_n(|\vec{p}+\vec{q}|)$ where $\vec{p}$ is now the momentum of the target proton inside the nucleus. Furthermore, in the nucleus the protons and neutrons are not free and their momenta are constrained to satisfy the Pauli principle, i.e., ${p_{p}}<{p_{F_{p}}}$ and ${p}_{n}(=|{\vec p}+{\vec q}|) > p_{F_{n}}$, where $p_{F_p}$ and $p_{F_n}$ are the local Fermi momenta of protons and neutrons at the interaction point in the nucleus and are given by $p_{F_p}=\left[3\pi^2\rho_p(r)\right]^\frac{1}{3}$ and $p_{F_n}=\left[3\pi^2\rho_n(r)\right]^\frac{1}{3}$, $\rho_p(r)$ and $\rho_n(r)$ are the proton and neutron nuclear densities. Moreover, in nuclei the threshold value of the reaction i.e. the Q-value of the reaction has to be taken into account.

To incorporate these modifications, the $\delta$ function in Eq.(5) i.e. $\delta[q_0+E_p-E_n]$ is modified to $\delta[q_0-Q+E_p(\vec{p})-E_n(\vec{p}+\vec{q})]$ and the factor
\begin{equation}
\int \frac{d\vec{p}}{(2\pi)^3}{n_p(\vec{p},\vec{r})}\frac{M_n M_p}{E_n E_p}\delta[q_0+E_p-E_n]
\end{equation}
occurring in Eq.(7) is replaced by $-(1/{\pi})$Im${{U_N}(q_0,\vec{q})}$, where ${{U_N}(q_0,\vec{q})}$ is the Lindhard function corresponding to the particle hole(ph) excitation shown in Fig.1 and is given by
\begin{equation}
{U_N}(q_0,\vec{q}) = {\int \frac{d\vec{p}}{(2\pi)^3}\frac{M_nM_p}{E_nE_p}\frac{n_p(p)\left[1-n_n(\vec p + \vec q) \right]}{q_0-Q+{E_p(p)}-{E_n(\vec p+\vec q)}+i\epsilon}}
\end{equation}
where the threshold value of the reaction, Q, in the present calculation is taken to be 13.6 MeV which corresponds to the lowest allowed Fermi transition. The expression for Im${U_N}$ is given in Ref.~\cite{singh}.
\begin{figure}
\includegraphics{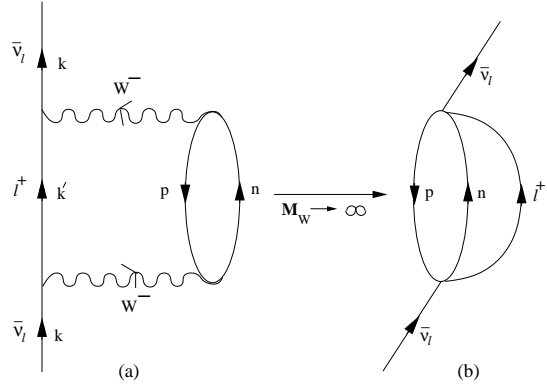}
\caption{Diagrammatic representation of the anti-neutrino self-energy diagram corresponding to the ph-excitation leading to ${\bar{\nu}}_\mu + p \rightarrow \mu^+ + n$ in nuclei. In the large mass limit of the IVB(i.e.$M_W\rightarrow \infty$) the diagram 1(a) is reduced to 1(b) which is used to calculate $|T|^2$ in Eq.(5).}
\end{figure}

With inclusion of these nuclear effects the cross section $\sigma_{A}(E_{\bar\nu})$ is written as
\begin{eqnarray}
\sigma_{A}(E_{\bar\nu})=-\frac{2{G_F}^2\cos^2{\theta_c}}{\pi}\int^{r_{max}}_{r_{min}} r^2 dr \int^{p_\mu^{max}}_{p_\mu^{min}}p_\mu^2dp_\mu
\int_{-1}^1d(cos\theta)\frac{1}{E_{{\bar{\nu}}_\mu} E_\mu} L_{\mu\nu}J^{\mu\nu} Im{U_N}[E_{{\bar{\nu}}_\mu} - E_\mu - Q, \vec{q}].
\end{eqnarray}
In the nucleus the strength of the electroweak coupling may change from their free nucleon values due to the presence of strongly interacting nucleons. Conservation of Vector Current (CVC) forbids any change in the charge coupling while magnetic and axial vector couplings are likely to change from their free nucleon values. These changes are calculated by considering the interaction of ph excitations in nuclear medium in Random Phase Approximation (RPA) as shown in Fig.2. The diagram shown in Fig.2 simulates the effects of the strongly interacting nuclear medium at the weak vertex. The ph-ph interaction is shown by the wavy line in Fig.2 and is described by the $\pi$ and $\rho$ exchanges modulated by the effect of short range correlations.

The weak nucleon current described by Eq.(4) gives, in non-relativistic limit, terms like $F_A \vec{\sigma}\tau_+$ and $i F_2 \frac{\vec{\sigma}\times \vec{q}}{2M}\tau_+$ which generate spin-isospin transitions in nuclei. While the term $i F_2 \frac{\vec{\sigma}\times \vec{q}}{2M}\tau_+$ couples to the transverse excitations, the term $F_A \vec{\sigma}\tau_+$ couples to the transverse as well as longitudinal channels. These channels produce different RPA responses in the longitudinal and transverse channels when the diagrams of Fig.(2) are summed over. For example, considering the renormalization of the axial vector term of the hadronic current in Eq.(4), the non-relativistic reduction of the axial vector term is written as
\begin{eqnarray}
\bar{u}(p^\prime)F_A\gamma^\mu\gamma^5u(p)=F_A(q^2)\left[\bar{u}(p^\prime)\gamma^0\gamma^5u(p),~\bar{u}(p^\prime)\gamma^i\gamma^5u(p)\right]\nonumber\\
=F_A(q^2)\left[\frac{\vec\sigma\cdot(\vec p+\vec {p^\prime})}{2E},~~~\left(\vec \sigma -\frac{\sigma^i(\vec \sigma\cdot \vec p)(\vec \sigma\cdot \vec {p^\prime})}{4E^2}\right)\right]~~
\end{eqnarray}
In leading order it is proportional to $F_A(q^2)\sigma^i$. One of the contributions of this term to the hadronic tensor $J^{ij}$ in the medium is proportional to $F^2_A(q^2)\delta_{ij}\mbox{Im}{U_N}$ which is split between the longitudinal and transverse components as:
\begin{equation}
F^2_A(q^2)\delta_{ij}\mbox{Im}{U_N}\rightarrow F^2_A(q^2)\left[{{\hat{q_i}}{\hat{q_j}}}+(\delta_{ij}-{{\hat{q_i}}{\hat{q_j}}})\right]\mbox{Im}{U_N}
\end{equation}
The RPA response of this term after summing the higher order diagrams like Fig.2 is modified and is given by $J_{RPA}^{ij}$:
\begin{equation}
J^{ij}\rightarrow J_{RPA}^{ij}= F^2_A(q^2)\mbox{Im}{U_N}\left[\frac{{{\hat{q_i}}{\hat{q_j}}}}{|1-U_NV_l|^2}+\frac{\delta_{ij}-{{\hat{q_i}}{\hat{q_j}}}}{|1-U_NV_t|^2}\right]
\end{equation}
where $V_l$ and $V_t$ are the longitudinal and transverse part of the nucleon-nucleon potential calculated with $\pi$ and $\rho$ exchanges and given by 
\begin{figure}
\includegraphics{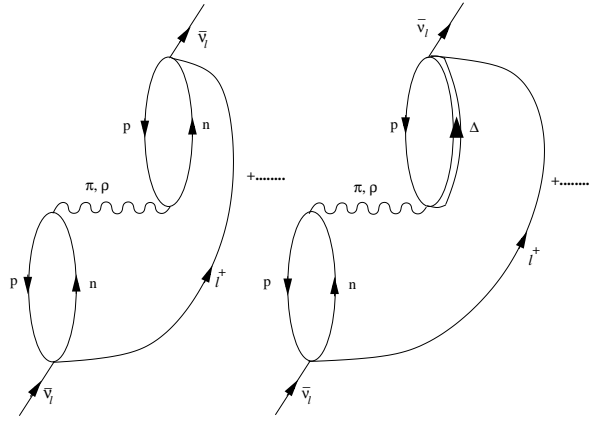}
\caption{ Many body Feynman diagrams (drawn in the limit $M_W\rightarrow \infty$) accounting for the medium polarization effects contributing to the process ${\bar{\nu}}_\mu + p \rightarrow \mu^+ + n$.}
\end{figure}
\begin{eqnarray}
V_l(q) = \frac{f^2}{m_\pi^2}\left[\frac{q^2}{-q^2+m_\pi^2}{\left(\frac{\Lambda_\pi^2-m_\pi^2}{\Lambda_\pi^2-q^2}\right)^2}+g^\prime\right],~~~~~V_t(q) = \frac{f^2}{m_\pi^2}\left[\frac{q^2}{-q^2+m^2_\rho}{C_\rho}{\left(\frac{{\Lambda_\rho}^2-m^2_\rho}{{\Lambda_\rho}^2-q^2}\right)^2}+g^\prime\right]
\end{eqnarray}
$\Lambda_\pi=1.3 GeV$, $C_\rho=2$, $\Lambda_\rho=2.5GeV$, $m_\pi$ and $m_\rho$ are the pion and $\rho$ masses, and $g^\prime$ is the Landau-Migdal parameter taken to be $0.7$ which has been used quite successfully to explain many electromagnetic and weak processes in nuclei~\cite{mukh}-\cite{gil}. This modified tensor $J^{ij}_{RPA}$ when contracted with the leptonic tensor $L_{ij}$ gives the contribution of the $F^2_A$ term to the differential cross sections including RPA correlations. 

The effect of the $\Delta$ degrees of freedom in nuclear medium is included in the calculation of the RPA response by considering the effect of ph-$\Delta$h and $\Delta$h-$\Delta$h excitations as shown in Fig.2. This is done by replacing $U_N$ by $U=U_N+U_\Delta$, where $U_\Delta$ is the Lindhard function for $\Delta$h excitation in the medium and the expressions for $U_N$ and $U_\Delta$ are taken from Ref.~\cite{oset1}. The different couplings of $N$ and $\Delta$ are incorporated in $U_N$ and $U_\Delta$ and then the same interaction strengths $V_l$ and $V_t$ are used to calculate the RPA response which has been discussed in some detail in Refs.~\cite{singh} and \cite{nieves}.

The treatment of Coulomb distortion of the produced muon in the Coulomb field of the final nucleus in the local density approximation is done by modifying the energy of the muon in the Coulomb field of the final nucleus:
\[ E_{eff} = E_\mu + V_c(r), \]
where 
\begin{equation}
V_c(r)=-Z_fZ^\prime\alpha 4\pi\left(\frac{1}{r}\int_0^r\frac{\rho_p(r^\prime)}{Z_f}{r^\prime}^2dr^\prime + \int_r^\infty\frac{\rho_p(r^\prime)}{Z_f}{r^\prime}dr^\prime\right)
\end{equation}
where $Z^\prime$ is 1 for $\mu^+$ and $\rho_p(r^\prime)$ is the proton density in the final nucleus.

Thus, in the presence of nuclear medium effects, the total cross section $\sigma_{A}(E_{\bar\nu})$, with the inclusion of Coulomb distortion effects taken into account is written as
\begin{eqnarray}
\sigma_{A}(E_{\bar\nu})=-\frac{2{G_F}^2\cos^2{\theta_c}}{\pi}\int^{r_{max}}_{r_{min}} r^2 dr \int^{p_\mu^{max}}_{p_\mu^{min}}{p_\mu}^2dp_\mu \int_{-1}^1d(cos\theta)~\frac{1}{E_{{\bar{\nu}}_\mu} E_\mu} L_{\mu\nu}{J^{\mu\nu}_{RPA}} Im{U_N}.
\end{eqnarray}
where 
\begin{eqnarray}
Im{U_N}&=&Im{U_N}[E_{{\bar{\nu}}_\mu} - E_\mu - Q-V_c(r), \vec{q}]
\end{eqnarray}

\section{Charged current inelastic reactions}
The inelastic production process of leptons is the process in which the production of leptons is accompanied by one or more pions. 
In the intermediate energy region of about 1~GeV the anti-neutrino induced reactions on a nucleon for lepton production is dominated by the $\Delta$ excitation which subsequently decays into a pion and a nucleon through the following reactions:
\begin{eqnarray}
\bar\nu_\mu(k)+ n(p)&\rightarrow& \mu^{+}(k^\prime)+\Delta^{-}(p^\prime)\\
       &&~~~~~~~~~~~~ \searrow n + \pi^-  \nonumber                       
\end{eqnarray}
\begin{eqnarray}
\bar\nu_\mu(k)+ p(p)&\rightarrow& \mu^{+}(k^\prime)+ \Delta^{0}(p^\prime)\\
           &&~~~~~~~~~~~~~~                     \searrow p + \pi^-\nonumber\\
           &&~~~~~~~~~~~~~~                       \searrow n  + \pi^0\nonumber
\end{eqnarray}

In this model of the $\Delta$ dominance the anti-neutrino induced charged current lepton production accompanied by a pion is calculated using the Lagrangian in the standard model of electroweak interactions given by Eq.(3), where the leptonic current is given by Eq.(4) and the hadronic current ${J^{\mu}(x)}=V^\mu(x)+ A^\mu(x)$.  

In this case, the hadronic current $J^{\mu}$ for the $\Delta$ excitation from neutron target is given by 
\[J^\mu=\bar{\psi}_\alpha(p^\prime)O^{\alpha\mu}{\psi}(p),\] where ${\psi_\alpha}(p^\prime)$ and ${\psi}(p)$ are the
Rarita Schwinger and Dirac spinors for the $\Delta$ and the nucleon of
momenta $p^\prime$ and $p$ respectively, and $O^{\alpha\mu}=O^{\alpha\mu}_V + O^{\alpha\mu}_A$. The operators $O^{\alpha\mu}_V$ and $O^{\alpha\mu}_A$ are given by:
\begin{eqnarray}
O^{\alpha\mu}_V&=&\left(\frac{C^V_{3}(q^2)}{M}(g^{\alpha\mu}{\not q}-q^\alpha{\gamma^\mu})+\frac{C^V_{4}(q^2)}{M^2}(g^{\alpha\mu}q\cdot{p^\prime}-q^\alpha{p^{\prime\mu}})\right.\nonumber\\
&+&\left.\frac{C^V_5(q^2)}{M^2}(g^{\alpha\mu}q\cdot p-q^\alpha{p^\mu})+\frac{C^V_6(q^2)}{M^2}q^\alpha q^\mu\right)\gamma_5
\end{eqnarray}
and
\begin{eqnarray}
O^{\alpha\mu}_A=\left(\frac{C^A_{3}(q^2)}{M}(g^{\alpha\mu}{\not q}-q^\alpha{\gamma^\mu})+\frac{C^A_{4}(q^2)}{M^2}(g^{\alpha\mu}q\cdot{p^\prime}-q^\alpha{p^{\prime\mu}})+C^A_{5}(q^2)g^{\alpha\mu}+\frac{C^A_6(q^2)}{M^2}q^\alpha q^\mu\right)
\end{eqnarray}
A similar expression for $J^\mu$ is used for the $\Delta^0$ excitation from the
proton target. Here $q(=p^\prime-p=k-k^\prime)$
is the momentum transfer, $Q^2$(=~-$q^2$) is the momentum transfer square and M is the mass of the nucleon. $C^V_i$(i=3-6)
are the vector and $C^A_i$(i=3-6) are the axial vector transition form
factors which have been taken from Ref.~\cite{paschos2} to be:
\begin{eqnarray}
C_i^V(q^2)&=&C_i^V(0)~\left(1-\frac{q^2}{M_V^2}\right)^{-2}~{D_i}; ~~i=3,4,5.\nonumber\\
{\it D_i}&=&\left(1-\frac{q^2}{4M_V^2}\right)^{-1}~for ~~i=3,4,~~~{\it D_i}=\left(1-\frac{q^2}{0.776M_V^2}\right)^{-1};~~i=5.
\end{eqnarray}
\begin{eqnarray}
C_i^A(q^2)=C_i^A(0)~~\left(1-\frac{q^2}{M_A^2}\right)^{-2}~{\it D_i};~~i=3,4,5.~~~{\it D_i}=\left(1-\frac{q^2}{3M_A^2}\right)^{-1} 
\end{eqnarray}
$C_6^V(q^2)$ and $C_6^A(q^2)$ are determined using conserved vector current (CVC) and partially conserved axial vector current (PCAC) hypothesis to be $C_6^V(q^2)=0$ and $C_6^A(q^2)=\frac{C_5^A(q^2)}{M^2-{m_\pi}^2}$. $M_A$(=1.05GeV) and $M_V$(=0.84GeV) are the axial vector and vector dipole masses, and $m_\pi$ is the pion mass. The values of $C_i^V(q^2)$ and $C_i^A(q^2)$ at $q^2$=0 are taken to be $C_3^V(0)$=2.13, $C_4^V(0)$=-1.51, $C_5^V(0)$=0.48, $C_3^A(0)$=0.0, $C_4^A(0)$=-0.25 and $C_5^A(0)$=1.2.

The differential scattering cross section is given by
\begin{equation}
\frac{d^2\sigma}{dE_{k^\prime}d\Omega_{k^\prime}}=\frac{1}{64\pi^3}\frac{1}{MM_\Delta}\frac{|{\bf
    k}^\prime|}{E_k}\frac{\frac{\Gamma(W)}{2}}{(W-M_\Delta)^2+\frac{\Gamma^2(W)}{4.}}{|{T}|^2}
\end{equation}
where $\Gamma$ is the delta decay width, $W$ is the center of mass energy i.e. $W=\sqrt{(p+q)^2}$ and ${|{T}|^2}= \frac{{G_F}^2\cos^2{\theta_c}}{2}L_{\mu\nu} J^{\mu\nu}$, with $L_{\mu\nu}$ given by Eq.(6) and $J^{\mu\nu}=\bar{\Sigma}\Sigma J^{\mu\dagger} J^\nu$ which is calculated with the use of spin $\frac{3}{2}$ projection operator $P^{\mu\nu}$ defined as \[P^{\mu\nu}=\sum_{spins}\psi^\mu {\bar{\psi^\nu}}\] and is given by:
\begin{equation}
P^{\mu\nu}=-\frac{\not{p^\prime}+M_\Delta}{2M_\Delta}\left(g^{\mu\nu}-\frac{2}{3}\frac{p^{\prime\mu} p^{\prime\nu}}{M_\Delta^2}+\frac{1}{3}\frac{p^{\prime\mu} \gamma^\nu-p^{\prime\nu} \gamma_\mu}{M_\Delta}-\frac{1}{3}\gamma^\mu\gamma^\nu\right)
\end{equation}
In Eq.(24), the delta decay width $\Gamma$ is taken to be an energy dependent P-wave decay width given by~\cite{a4}:
\begin{eqnarray}
\Gamma(W)=\frac{1}{6\pi}\left(\frac{f_{\pi N\Delta}}{m_\pi}\right)^2\frac{M}{W}|{{\bf q}_{cm}|^3}\Theta(W-M-m_\pi),
\end{eqnarray}
where 
\[|{\bf q}_{cm}|=\frac{\sqrt{(W^2-m_\pi^2-M^2)^2-4m_\pi^2M^2}}{2W}\]
and $M$ is the mass of nucleon. The step function $\Theta$ denotes the fact that the width is zero for the invariant masses below the $N\pi$ threshold, ${|\bf q_{cm}|}$ is the pion momentum in the rest frame of the resonance.

When the reactions given by Eq.(18) or (19) take place in the nucleus, the
anti-neutrino interacts with a nucleon moving inside the nucleus of density
$\rho(r)$ with its corresponding momentum $\vec{p}$ constrained to be
below its Fermi momentum
$p_{F_{n,p}}(r)=\left[3\pi^2\rho_{n,p}(r)\right]^\frac{1}{3}$, where
$\rho_n(r)$ and $\rho_p(r)$ are the neutron and proton nuclear
densities. In the local density approximation, the differential
scattering cross section for a lepton production accompanied by a pion from the neutron target is written as
\begin{eqnarray}
\frac{d^2\sigma}{dE_{k^\prime}d\Omega_{k^\prime}}=\frac{1}{64\pi^3}\int
d{\bf r}\rho_n({\bf r})\frac{|{\bf k}^\prime|}{E_k}\frac{1}{MM_\Delta}\frac{\frac{\Gamma(W)}{2}}{(W-M_\Delta)^2+\frac{\Gamma^2(W)}{4.}}|{T}|^2  
\end{eqnarray}
However, in nuclear medium the properties of $\Delta$ like its
mass and decay width $\Gamma$ to be used in Eq.(27) are modified due
to the nuclear effects. These are mainly due to the following processes.

(i) In nuclear medium $\Delta$s decay mainly through the $\Delta \rightarrow N\pi$ channel. The final nucleons have to be above the Fermi momentum $p_F$ of the nucleon in the nucleus thus inhibiting the decay as compared to the free decay of the $\Delta$ described by $\Gamma$ in Eq.(27). This leads to a modification in the decay width of delta which has been studied by many authors~\cite{a4},~\cite{oset3}-\cite{hofman}. We take the value given by Oset et al.~\cite{a4} and write the modified delta decay width $\tilde\Gamma$ as
\begin{equation}
\tilde\Gamma=\Gamma \times F(p_{F},E_{\Delta},k_{\Delta})
\end{equation}
 where $F(p_{F},E_{\Delta},k_{\Delta})$ is the Pauli correction factor given by~\cite{a4}:
\begin{equation}
F(p_{F},E_{\Delta},k_{\Delta})= \frac{k_{\Delta}|{{\bf q}_{cm}}|+E_{\Delta}{E^\prime_p}_{cm}-E_{F}{W}}{2k_{\Delta}|{\bf q^\prime}_{cm}|} 
\end{equation}
 $E_F=\sqrt{M^2+p_F^2}$, $k_{\Delta}$ is the $\Delta$ momentum and  $E_\Delta=\sqrt{W+k_\Delta^2}$. 

(ii) In nuclear medium there are additional decay channels open
due to two and three body absorption processes like $\Delta N
\rightarrow N N$ and $\Delta N N\rightarrow N N N$ through which
$\Delta$ disappears in nuclear medium without producing a pion,
while a two body $\Delta$ absorption process like $\Delta N
\rightarrow \pi N N$ gives rise to some more pions. These nuclear
medium effects on the $\Delta$ propagation are included by describing
the mass and the decay width in terms of the self energy of
$\Delta$~\cite{a4}. The real part of the $\Delta$ self energy gives modification in the mass and the imaginary part of the $\Delta$ self energy gives modification in the decay width of $\Delta$ inside the nuclear medium. The expressions for the real and imaginary part of the $\Delta$ self energy are taken from Oset et al.~\cite{a4}:
\begin{eqnarray}
Re{\Sigma}_{\Delta}&=&40 \frac{\rho}{\rho_{0}}MeV ~~and \nonumber\\
-Im{{\Sigma}_{\Delta}}&=&C_{Q}\left (\frac{\rho}{{\rho}_{0}}\right )^{\alpha}+C_{A2}\left (\frac{\rho}{{\rho}_{0}}\right )^{\beta}+C_{A3}\left (\frac{\rho}{{\rho}_{0}}\right )^{\gamma}~~~~
\end{eqnarray}
In the above equation $C_{Q}$ accounts for the $\Delta N  \rightarrow
\pi N N$ process, $C_{A2}$ for the two-body absorption process $\Delta
N \rightarrow N N$ and $C_{A3}$ for the three-body absorption process $\Delta N N\rightarrow N N N$. The coefficients $C_{Q}$, $C_{A2}$, $C_{A3}$ and $\alpha$, $\beta$ and $\gamma$ are taken from Ref.~\cite{a4}.

These considerations lead to the following modifications in the width $\tilde\Gamma$ and mass $M_\Delta$ of the $\Delta$ resonance. 
\begin{equation}
\frac{\Gamma}{2}\rightarrow\frac{\tilde\Gamma}{2} - Im\Sigma_\Delta~~\text{and}~~
M_\Delta\rightarrow\tilde{M}_\Delta= M_\Delta + Re\Sigma_\Delta.
\end{equation}
With these modifications the differential scattering cross section described by Eq.(27) on the neutron target modifies to 
\begin{eqnarray}
\frac{d^2\sigma}{dE_{k^\prime}d\Omega_{k^\prime}}=\frac{1}{64\pi^3}\int
d{\bf r}\rho_n({\bf r})\frac{|{\bf k}^\prime|}{E_k}\frac{1}{MM_\Delta}\frac{\frac{\tilde\Gamma}{2}-Im\Sigma_\Delta}{(W- \tilde{M}_\Delta)^2+(\frac{\tilde\Gamma}{2.}-Im\Sigma_\Delta)^2}|{T}|^2
\end{eqnarray}
For the lepton production from proton target, $\rho_n({\bf r})$ in the
above expression is replaced by $\frac{1}{3}\rho_p({\bf r})$. Therefore, the total scattering cross section for the anti-neutrino induced charged current lepton production process in the nucleus is given by
\begin{eqnarray}
\sigma =\frac{1}{64\pi^3}\int \int {d{\bf r}}\frac{d\bf{k^\prime}}{E_k E_{k^\prime}}\frac{1}{MM_\Delta}\frac{\frac{\tilde\Gamma}{2}-Im\Sigma_\Delta}{(W- \tilde{M}_\Delta)^2+(\frac{\tilde\Gamma}{2.}-Im\Sigma_\Delta)^2}\left[\rho_n({\bf r})+\frac{1}{3}\rho_p({\bf r})\right]|{T}|^2
\end{eqnarray}
\subsection{Inelastic production of leptons accompanied by a pion}
The reactions given in Eqs.(18) and (19) produce $\pi^-$ and $\pi^0$ in the nucleus. The target nucleus can stay in the ground state giving all the transferred energy in the reaction to the outgoing pion leading to the coherent production of pions or can be excited and/or broken up leading to the incoherent production of pions. In this section we discuss the inelastic charged current lepton production accompanied by a $\pi^-$ or $\pi^0$. The inelastic coherent production of leptons is discussed in section-B. 

In the nucleus the incoherent production of lepton takes place through the production and decay of $\Delta^-$ on neutron targets and the production and decay of $\Delta^0$ on proton targets. Some of the $\Delta$s are absorbed through two body and three body absorption processes and do not lead to pion production. These are described by $C_{A2}$ and $C_{A3}$ terms in the expression for $Im{{\Sigma}_{\Delta}}$ given in Eq.(30) and do not contribute to the lepton production accompanied by pions. Thus only $C_Q$ term in the expression for Im$\Sigma_\Delta$ in Eq.(30) contributes to the lepton production accompanied by a pion.

Taking into consideration the appropriate factors corresponding to the $\Delta$ production on neutron and proton targets in the nucleus, we get the following expression for the inelastic production of leptons accompanied by a pion (i.e. $\pi^-$ or $\pi^0$).
\begin{eqnarray}
\sigma=\frac{1}{64\pi^3}\int \int {d{\bf r}}\frac{d\bf{k^\prime}}{E_k E_{k^\prime}}\frac{1}{MM_\Delta}\frac{\frac{\tilde\Gamma}{2}+C_{Q}\left (\frac{\rho}{{\rho}_{0}}\right )^{\alpha}}{(W- M_\Delta-Re\Sigma_\Delta)^2+(\frac{\tilde\Gamma}{2.}-Im\Sigma_\Delta)^2}\left[\rho_n({\bf r})+\frac{1}{3}\rho_p({\bf r})\right]|{T}|^2
\end{eqnarray}
The pions produced in these processes inside the nucleus may
re-scatter or may produce more pions or may get absorbed while coming
out from the final nucleus. We have taken the results of Vicente
Vacas~\cite{a5} for the final state interaction of
pions which is calculated in an eikonal approximation using probabilities per unit
length as the basic input. In this approximation, a pion of given
momentum and charge is moved along the z-direction with a
random impact parameter ${\bf b}$, with $|{\bf b}|<R$, where R is the
nuclear radius which is taken to be a point where nuclear density
$\rho(R)$ falls to ${10}^{-3}\rho_0$, where $\rho_0$ is the central
density. To start with, the pion is placed at a point $({\bf b},
z_{in})$, where $z_{in}=-\sqrt{R^2-|{\bf b}|^2}$ and then it is moved in
small steps $\delta l$ along the z-direction until it comes out of the
nucleus or interact. If $P(p_\pi,r,\lambda)$ is the probability per
unit length at the point r of a pion of momentum ${\bf p}_\pi$ and
charge $\lambda$, then $P\delta l <<1$. A random number x is generated
such that $x\in [0,1]$ and if $x > P\delta l$, then it is assumed
that pion has not interacted while traveling a distance
$\delta l$, however, if $x < P\delta l$ then the
pion has interacted and depending upon the weight factor of
each channel given by its cross section it is decided that whether the
interaction was
 quasielastic, charge exchange reaction, pion production or pion absorption~\cite{a5}. For example, for the quasielastic scattering
\begin{equation*}
P_{N(\pi^\lambda,\pi^{\lambda^\prime})N^\prime}=\sigma_{N(\pi^\lambda,\pi^{\lambda^\prime})N^\prime}\times
\rho_N
\end{equation*}
where N is a nucleon, $\rho_N$ is its density and $\sigma$ is the
elementary cross section for the reaction $\pi^\lambda +N \rightarrow
\pi^{\lambda^\prime} + N^\prime$ obtained from the phase shift
analysis. 

For a pion to be absorbed, $P$ is expressed in terms of the imaginary part
of the pion self energy $\Pi$
i.e. $P_{abs}=-\frac{Im\Pi_{abs}(p_\pi)}{p_\pi}$, where the self energy $\Pi$ is
related to the pion optical potential $V$~\cite{a5}.
\subsection{Coherent Pion Production}
The neutrino induced coherent production of pion has been calculated earlier in this
model~\cite{a6}, where $\Delta$ resonance excitations and their decays
are such that the nucleus stays in the ground state. We apply this model to calculate the coherent pion production induced by charged current anti-neutrino interaction inside the nucleus. 
\begin{figure}[h]
\includegraphics{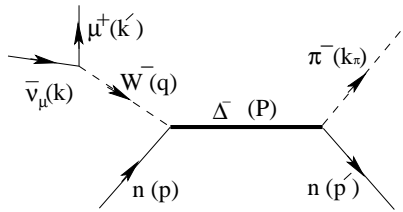}
\caption{Feynman diagram}
\end{figure}\\
The amplitude for charged current 1$\pi^-$ production from the neutron is written using the Feynman diagram shown in Fig.3 and is given by
\begin{equation}
{\cal A}~=~\frac{G_F}{\sqrt{2}}~\cos\theta_C~l^\mu~{\cal J}_\mu~{\cal F}({\bf q}-{\bf k_\pi})
\end{equation}
where $l^\mu$ is the leptonic current given in Eq.(4) and the hadronic current ${\cal J}_\mu$ is given by
\begin{eqnarray}
{\cal J}_\mu=\frac{f_{{\pi} N {\Delta}}}{m_{\pi}}\sum_s{\bar {\Psi}}^s(p^\prime)\left[k_{\pi\sigma}\Lambda^{\sigma \lambda}{\cal{O}}_{\lambda \mu}\right]\Psi^s(p)
\end{eqnarray}
where $\Lambda^{\sigma \lambda}$ is the relativistic $\Delta$ propagator given by 
\begin{eqnarray}
{\Lambda}^{\sigma\lambda}={\frac{{\not {P}}+M_\Delta}{P^2-M^{2}_{\Delta}+i\Gamma M_\Delta}}\left[g^{\sigma \lambda}-\frac{1}{3}\gamma^\sigma\gamma^\lambda-\frac{2}{3M^2_\Delta}P^\sigma P^\lambda+\frac{(P^\sigma \gamma^\lambda-\gamma^\sigma P^\lambda)}{3M_\Delta}\right]
\end{eqnarray}
and ${\cal{O}}^{\sigma\lambda}$ is the weak N-$\Delta$ transition vertex given as the sum of vector and axial part using Eqs.(20) and (21). The nuclear form factor ${\cal F}({\bf q}-{\bf k_\pi})$ in Eq.(35) is given as
\begin{equation}
{\cal F}({\bf q}-{\bf k}_\pi)=\int d^{3}{\bf r}~\rho({\bf r})~e^{-i({\bf q}-{\bf k}_\pi).{\bf r}}
\end{equation}
with $\rho({\bf r})$ as the nuclear matter density as a function of nucleon relative coordinates. For production from nuclear targets, it is the linear combination of proton and neutron densities incorporating the isospin factors for charged pion production from proton and neutron targets corresponding to $W^-$ exchange diagram. It is written as
\begin{equation}
{\cal F}({\bf q}-{\bf k}_\pi)=\int d^{3}{\bf r} \left[\frac{1}{3}{\rho_p ({\bf r})}+{\rho_n ({\bf r})}\right]e^{-i({\bf q}-{\bf k}_\pi).{\bf r}}
\end{equation} 

Using these expressions the following form of the differential cross section for lepton production is obtained:
\begin{equation}
\frac{d^5\sigma}{d\Omega_\pi d\Omega_{{\bar\nu}\mu}dE_\mu}~=~\frac{1}{8}\frac{1}{(2\pi)^5}\frac{|~{\bf k^\prime}|~|~{\bf k_\pi}|}{E_{\bar\nu}}~{\cal R}~{\bar{\sum}}\sum|{\cal A}|^{2}
\end{equation}
where
\begin{equation}
{\cal R}=\left[\frac{M~|{\bf k_\pi}|}{E_{p^\prime}|{\bf k_\pi}|+E_\pi(|{\bf k_\pi}|-|{\bf q}|\cos\theta_\pi)}\right]
\end{equation}
is a kinematical factor incorporating the recoil effects, which is very close to unity for low $Q^2(=-q^2)$, relevant for the coherent reactions, and $|{\cal A}|^{2}$ is obtained by squaring the terms given in Eq.(35).

The nuclear medium effects due to renormalization of $\Delta$ properties in the nuclear medium have been treated in the same manner as discussed in section-III. Accordingly the $\Delta$ propagator $\Lambda^{\sigma \lambda}$ in ${\cal J}_\mu$ given by Eq.(37) is modified due to the modifications in mass and width of the $\Delta$ in the nuclear medium given by Eq.(31). However, the final state interaction of the pions with the residual nucleus has been treated in a different way. The final state interaction in coherent production of pions is taken into account by replacing the plane wave pion by a distorted wave pion. The distortion of the pion is calculated in the eikonal approximation~\cite{oset6} in which the distorted pion wave function is written as:
\begin{equation}
e^{i({\bf q}-{\bf k}_\pi)\cdot{\bf r}}\rightarrow \mbox{exp}\left[i({\bf{q-k_\pi}})\cdot{\bf{r}}-\frac{i}{v}\int_{-\infty}^z V_{opt}({\bf b}, z^\prime)dz^\prime\right]
\end{equation}
where ${\bf r}=({\bf b}, z)$, ${\bf q}$ and ${\bf k}_\pi$ are the momentum transfer and the pion momentum, respectively. The pion optical potential $V_{opt}$ is related with the pion self-energy $\Pi$ as $\Pi=2\omega~V_{opt}$, where $\omega$ is the energy of the pion and $|{\bf{v}}|={|{\bf{k_\pi}}|}/{\omega}$. The pion self-energy is calculated in local density approximation of the $\Delta$-hole model and is given as~\cite{a4}:
\begin{equation}
\Pi(\rho({\bf b}, z^\prime))=\frac{4}{9}\left(\frac{f_{\pi N\Delta}}{m_\pi}\right)^2\frac{M^2}{\bar s}|{\bf k_\pi}|^2~\rho({\bf b}, z^\prime)~G_{\Delta h}({\bar s}, \rho)
\end{equation}
where $\bar{s}$ is the center of mass energy in the $\Delta$ decay averaged over the Fermi sea and $G_{\Delta h}({\bar s}, \rho)$ the $\Delta$-hole propagator given by~\cite{oset6}:
\begin{eqnarray}
G_{\Delta h}(s,\rho({\bf b}, z^\prime))=\frac{1}{\sqrt{{\bar s}}-M_\Delta+\frac{1}{2}i\tilde\Gamma({\bar s},\rho)-i\mbox{Im}\Sigma_\Delta({\bar s}, \rho)-\mbox{Re}\Sigma_\Delta({\bar s},\rho)}
\end{eqnarray}
When the pion absorption effect is taken into account the nuclear form factor ${\cal F}({\bf q}-{\bf k_\pi})$ modifies to $\tilde{\cal F}({\bf q}-{\bf k_\pi})$ given as:
\begin{eqnarray}
\tilde{\cal F}({\bf q}-{\bf k_\pi})=2\pi\int_0^\infty b~db\int_{-\infty}^\infty dz~\rho({\bf b}, z)\left[J_0(k_\pi^tb)~e^{i(|{\bf q}|-k_\pi^l)z}~e^{-if({\bf b}, z)}\right]
\end{eqnarray}
where 
\begin{equation}
f({\bf b}, z)=\int_z^{\infty} \frac{1}{2|{\bf{k}_\pi}|}{\Pi(\rho({\bf b}, z^\prime))}dz^\prime
\end{equation}
and the pion self-energy $\Pi$ is defined in Eq.(43).

These modifications lead to the following expression for the total scattering cross section 
\begin{equation}
\sigma=\frac{1}{8}\frac{1}{(2\pi)^5}\int d\Omega_\pi\int d\Omega_{{\bar\nu}\mu}\int dE_\mu\frac{|~{\bf k^\prime}|~|~{\bf k_\pi}|}{E_{\bar\nu}}~{\cal R}~{\bar{\sum}}\sum|{\tilde A}|^{2}
\end{equation}
where
\begin{eqnarray}
{\tilde A}&=&\frac{G_F}{\sqrt{2}}~\cos\theta_C~l^\mu~\tilde{\cal J}_\mu~\tilde{\cal F}({\bf q}-{\bf k_\pi})\nonumber\\
\tilde{\cal J}_\mu&=&\frac{f_{{\pi} N {\Delta}}}{m_{\pi}}\sum_s{\bar {\Psi}}^s(p)\left[k_{\pi\sigma}\tilde\Lambda^{\sigma \lambda}{\cal{O}}_{\lambda \mu}\right]\Psi^s(p)
\end{eqnarray}
where $\tilde\Lambda$ is the modified $\Delta$ propagator inside the nuclear medium.
\subsection{Incoherent production of 1$\pi^-$}
The incoherent lepton production accompanied by a $\pi^-$ is mainly given by the decay of $\Delta^-$ and $\Delta^0$ particles produced from neutron and proton targets in the nuclear medium through $\Delta^-\rightarrow n\pi^-$ and $\Delta^0\rightarrow p\pi^-$ processes given in Eqs.(18) and (19). These are mainly described by the modified decay width $\tilde\Gamma$ in the nuclear medium. However, there is an additional contribution to the pion production coming from the $C_{Q}$ term in the expression of $Im\Sigma_\Delta$ given in Eq.(30). Therefore, the total scattering cross section for the anti-neutrino induced charged current one $\pi^-$ production in the nucleus is given by
\begin{eqnarray}
\sigma=\frac{1}{64\pi^3}\int \int {d{\bf r}}\frac{d\bf{k^\prime}}{E_k E_{k^\prime}}\frac{1}{MM_\Delta}\frac{\frac{\tilde\Gamma}{2}+C_{Q}\left (\frac{\rho}{{\rho}_{0}}\right )^{\alpha}}{(W- M_\Delta-Re\Sigma_\Delta)^2+(\frac{\tilde\Gamma}{2.}-Im\Sigma_\Delta)^2}\left[\rho_n({\bf r})+\frac{1}{9}\rho_p({\bf r})\right]|{T}|^2
\end{eqnarray}
The final state interaction of these pions are treated in the same way as discussed in section-III(A).
\subsection{Quasielastic like production of leptons}
In a nuclear medium when an anti-neutrino interacts with a nucleon inside the nucleus, the $\Delta$ which is formed may disappear through two and three body absorption processes like $\Delta N
\rightarrow N N$ and $\Delta N N\rightarrow N N N$ and thus mimic a quasielastic reaction discussed in section-II. These $\Delta$ absorption processes are described by the $C_{A2}$ and $C_{A3}$ terms in the expression of Im$\Sigma_\Delta$ given in Eq.(30). To estimate the number of quasielastic like lepton events(without a pion) we write the expression for the total scattering cross section using Eq.(30) as
\begin{eqnarray}
\sigma=\frac{1}{64\pi^3}\int \int {d{\bf r}}\frac{d\bf{k^\prime}}{E_k E_{k^\prime}}\frac{1}{MM_\Delta}\frac{C_{A2}\left (\frac{\rho}{{\rho}_{0}}\right )^{\beta}+C_{A3}\left (\frac{\rho}{{\rho}_{0}}\right )^{\gamma}}{(W- M_\Delta-Re\Sigma_\Delta)^2+(\frac{\tilde\Gamma}{2.}-Im\Sigma_\Delta)^2}\left[\rho_n({\bf r})+\frac{1}{3}\rho_p({\bf r})\right]|{T}|^2
\end{eqnarray}
\section{Results and Discussion}
\subsection{Charged current quasielastic lepton production}
We present the numerical results for the total cross sections, $Q^2$ distributions and lepton angular distributions for the quasielastic charged current lepton production process ${\bar{\nu}}_\mu + ^{12}C \rightarrow \mu^+ + ^{12}B^*$. The calculations have been done using Eq.(16) with proton density $\rho_\text{p}(r)=\frac{Z}{A}\rho(r)$ and the neutron density $\rho_\text{n}(r)=\frac{A-Z}{A}\rho(r)$, where $\rho(r)$ is nuclear density taken as 3-parameter Fermi density given by:
\[\rho(r)=\rho_0\left(1+w\frac{r^2}{c^2}\right)/\left(1+exp\left(\frac{r-c}{z}\right)\right),\] the density parameters c=2.355fm, z=0.5224fm and w=-0.149 taken from Ref.~\cite{vries}. 
\begin{figure}[h]
\includegraphics{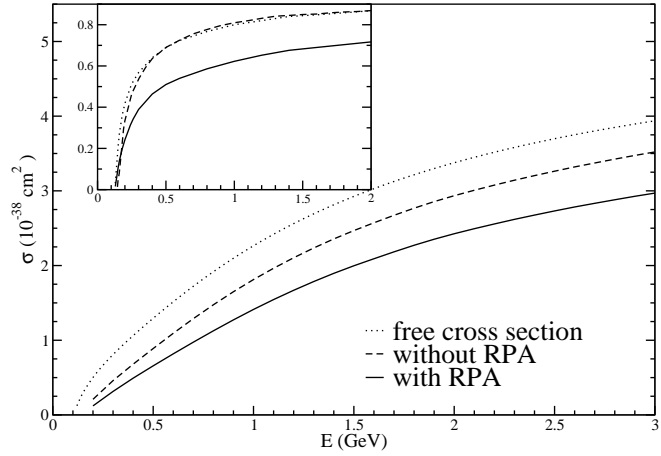}
\caption{Quasielastic charged current lepton production cross section induced by
 ${\bar\nu}_\mu$ on $^{12}$C target for the free protons and including nuclear medium effects with(without) RPA correlations. In the inset, ratio of the cross section per nucleon in $^{12}$C to the free proton cross section as a function of neutrino energy is shown. The solid line (dashed line) is the result for this ratio for the cross section calculated with nuclear medium effects with(without) RPA correlations. The dotted line is the prediction from the NUANCE Monte Carlo generator~\cite{a1}.}
\end{figure}

The numerical results for the total cross section $\sigma(E)$ vs E for anti-neutrino reactions on $^{12}C$ has been shown in Fig.4. The results have been presented for the cross section calculated for the free nucleon, with nuclear medium effects in local Fermi gas model with and without RPA correlations in nuclear medium. We find that with the incorporation of nuclear medium effects without RPA correlations the reduction in the cross section is around $35\%$ at $E_{{\bar\nu}_\mu}$=0.4GeV, $25\%$ at $E_{{\bar\nu}_\mu}$=0.8GeV, $18\%$ at $E_{{\bar\nu}_\mu}$=1.2GeV, $15\%$ at $E_{{\bar\nu}_\mu}$=1.6GeV and $12\%$ at $E_{{\bar\nu}_\mu}$=3.0GeV as compared with the free nucleon case. When RPA correlations are included the total reduction in the cross section is around $55\%$ at $E_{{\bar\nu}_\mu}$=0.4GeV, $40\%$ at $E_{{\bar\nu}_\mu}$=0.8GeV, $35\%$ at $E_{{\bar\nu}_\mu}$=1.2GeV, $30\%$ at $E_{{\bar\nu}_\mu}$=1.6GeV and $25\%$ at $E_{{\bar\nu}_\mu}$=3.0GeV. This reduction in the cross section is explicitly shown in the inset where we have presented the results for the ratio of the charged current quasielastic lepton production cross section to the cross section on free proton i.e. $\frac{1}{Z}\frac{\sigma(^{12}C)}{\sigma(free)}$ as a function of anti-neutrino energy. The results have been compared with the results obtained in the Fermi gas model used in the NUANCE Monte Carlo generator~\cite{nuance1} by the MiniBooNE collaboration~\cite{a1}. We find that the present results in the local Fermi gas model without RPA correlations are similar to the results used in the NUANCE generator, but when RPA effects are included the cross sections are reduced. 
\begin{figure}
\includegraphics{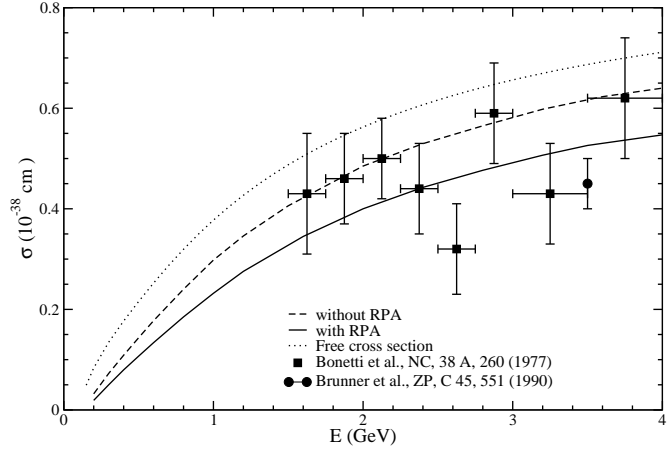}
\caption{Charged current quasielastic lepton production cross section per nucleon induced by
 ${\bar\nu}_\mu$ on Freon($CF_3Br$), for the free case and including nuclear medium effects with(without) RPA correlations. The experimental points are taken from Bonetti et al.~\cite{bon} and Brunner et al.~\cite{bru}.}
\end{figure}
\begin{figure}
\includegraphics{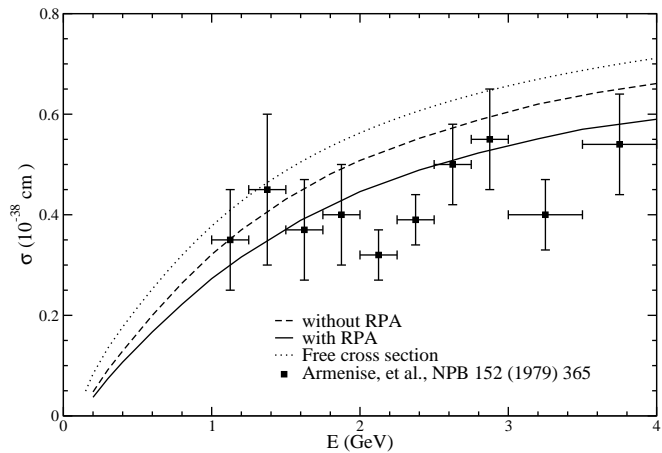}
\caption{Charged current quasielastic lepton production cross section per nucleon induced by
 ${\bar\nu}_\mu$ on Freon Propane($CF_3Br-C_3H_8$), for the free case and including nuclear medium effects with(without) RPA correlations. The experimental points are taken from Armenise et al.~\cite{armenise}.}
\end{figure}

In Figs. 5 and 6, we present our results for the total cross section obtained for the charged current quasielastic lepton production cross section induced by ${\bar\nu}_\mu$ on Freon($CF_3Br$) and Freon-Propane($CF_3Br-C_3H_8$) respectively. We have compared our results with the experimental results of Bonetti et al.~\cite{bon} and Brunner et al.~\cite{bru} in Freon($CF_3Br$) and in Freon-Propane by Armenise et al.\cite{armenise}. Quantitatively, we find that when RPA correlations are included $\chi^2$pdf are reduced to 0.5 from 1.7 for the Freon data and from 7.2 to 1.7 for the Freon-Propane data from $\chi^2$pdf calculated for the free case.

In Figs. 7 and 8, we have presented respectively the results for the $Q^2$-distribution i.e. $\langle\frac{d\sigma}{dQ^2}\rangle$ vs $Q^2$ and lepton angular distribution i.e. $\langle\frac{d\sigma}{dcos\theta}\rangle$ vs $cos\theta$, averaged over the MiniBooNE spectrum of anti-neutrino. The MiniBooNE spectrum is taken from Ref.~\cite{boone1}. The results of the differential cross sections have been presented for the free nucleon case as well as with nuclear medium effects. This will be useful in determining the axial vector form factor from the low energy anti-neutrino data obtained from MiniBooNE collaboration. We find that for $Q^2$ distribution the reduction in the differential cross section in the local Fermi gas model is around 30\% in the peak region of $Q^2$. When RPA effects are also taken into account the total reduction is around 50\% in this energy region. This reduction in the differential cross section decreases with the increase in $Q^2$, for example, at 0.2$GeV^2$ the total reduction is around 30\%. In the case of angular distribution, we find that at forward angles the reduction in the differential cross section when calculated in the local Fermi gas model is around 30-40\%. When RPA effects are also taken into account the total reduction is around 50-60\%. To quantitatively show the effect of RPA correlations on $Q^2$ and angular distributions, we have presented in the inset of these figures the ratio of differential cross sections calculated taking into account nuclear medium effects in local Fermi gas model with and without RPA correlations. 
\begin{figure}
\includegraphics{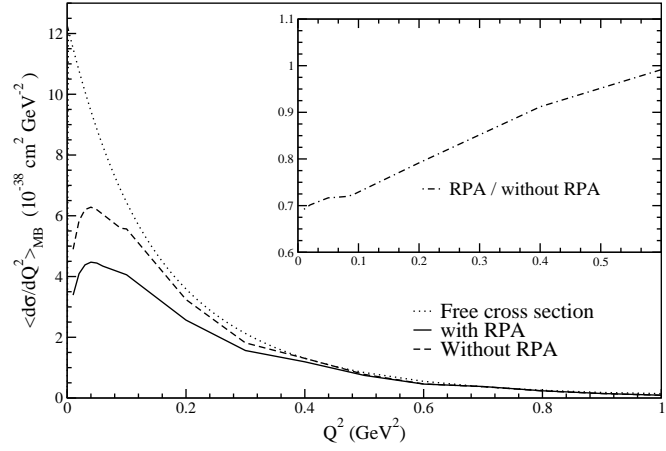}
\caption{$\langle\frac{d\sigma}{dQ^2}\rangle$ vs $Q^2$ for ${\bar\nu}_\mu$ on $^{12}$C target averaged over the MiniBooNE spectrum for the charged current quasielastic lepton production process. for the free case and with nuclear medium effects including RPA (without RPA). In the inset the ratio of $\langle\frac{d\sigma}{dQ^2}\rangle$ with and without RPA effects has been shown.}
\end{figure}
\begin{figure}
\includegraphics{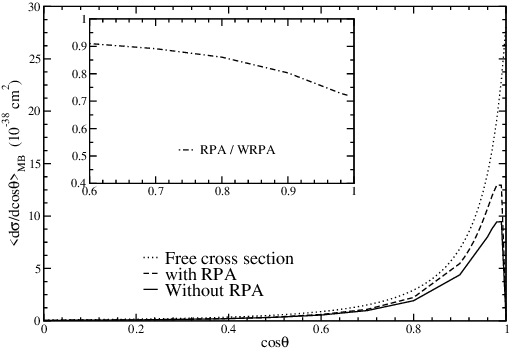}
\caption{$\langle\frac{d\sigma}{dcos\theta}\rangle$ vs $cos\theta$ for ${\bar\nu}_\mu$ on $^{12}$C target averaged over the MiniBooNE spectrum for the charged current quasielastic lepton production process, for the free case and with nuclear medium effects including RPA (without RPA). In the inset the ratio of $\langle\frac{d\sigma}{dQ^2}\rangle$ with and without RPA effects has been shown.}
\end{figure}
\subsection{ Charged current quasielastic like production of leptons}
In this section, we present the numerical results for the charged current anti-neutrino induced quasielastic like lepton production. These are the leptons produced through the inelastic process of $\Delta$ excitation in which, $\Delta$ is subsequently absorbed. This is purely a nuclear medium effect. These are discussed in section-III-D. The numerical calculations have been done by using Eq.(50) with N-$\Delta$ transition form factors given by Lalakulich et al.~\cite{paschos2}. In Fig.9, we present the results for the total scattering cross sections $\sigma(E)$ vs E on $^{12}C$ for the quasielastic lepton production cross section with nuclear medium effects including RPA, and the quasielastic like lepton production in the delta dominance model. We find that the quasielastic like lepton production contributes around 8\% at $E_{{\bar\nu}_\mu}$=1GeV and 12\% at $E_{{\bar\nu}_\mu}$=2-3GeV to the total quasielastic lepton production. In Figs.10 and 11, we present the results for the quasielastic lepton production cross section with nuclear medium effects including RPA, and the quasielastic like lepton production in the delta dominance model for Freon($CF_3Br$) and Freon-Propane($CF_3Br-C_3H_8$) respectively. These results are compared with the experimental results of Bonetti et al.~\cite{bon} and Brunner et al.~\cite{bru} in Freon($CF_3Br$) and in Freon-Propane by Armenise et al.\cite{armenise}. We find that when the total lepton production cross sections from these two contributions are taken into account and compared with the experimental data, $\chi^2$pdf is found to be 0.45 for the Freon data and 0.55 for the Freon-Propane data. Thus the inclusion of quasielastic like events improves the agreement with the experimental data for both the experiments currently available in literature.
\begin{figure}
\includegraphics{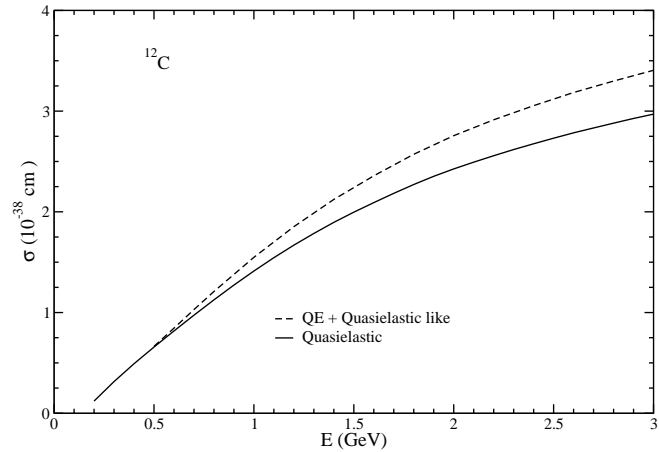}
\caption{Charged current lepton production cross section induced by
 ${\bar\nu}_\mu$ on $^{12}$C. The result shown by solid line is the quasielastic lepton production cross section with nuclear medium effects including RPA, and the result shown by dashed line also includes the quasielastic like lepton production cross section in the delta dominance model.}
\end{figure}
\begin{figure}
\includegraphics{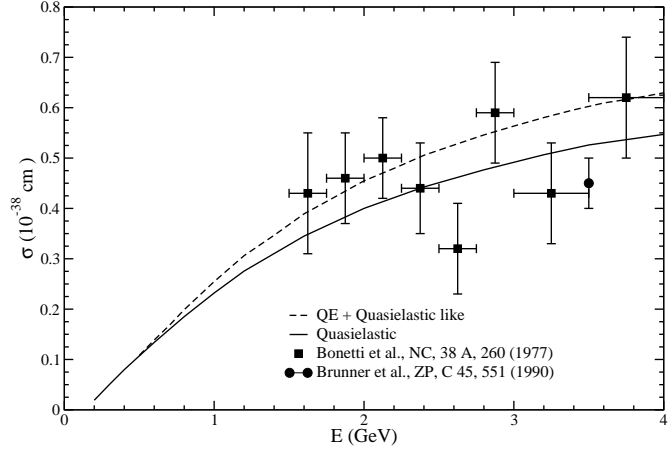}
\caption{Charged current lepton production cross section per nucleon induced by
 ${\bar\nu}_\mu$ on Freon($CF_3Br$). The result shown by solid line is the quasielastic lepton production cross section with nuclear medium effects including RPA, and the result shown by dashed line also includes the quasielastic like lepton production cross section in the delta dominance model. The experimental points are taken from Bonetti et al.~\cite{bon} and Brunner et al.~\cite{bru}.}
\end{figure}
\begin{figure}
\includegraphics{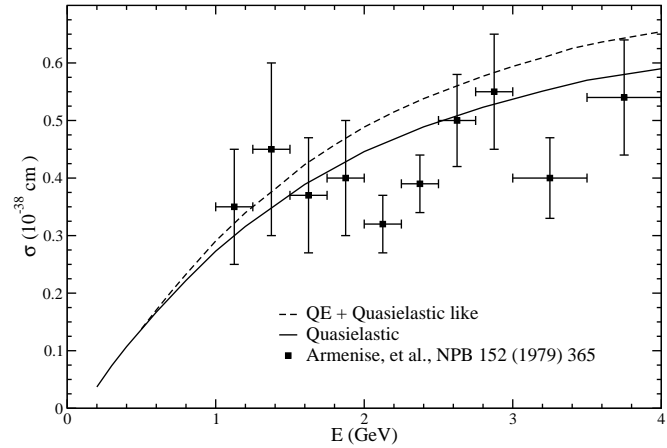}
\caption{Charged current lepton production cross section per nucleon induced by
 ${\bar\nu}_\mu$ on Freon Propane($CF_3Br-C_3H_8$). The result shown by solid line is the quasielastic lepton production cross section with nuclear medium effects including RPA, and the result shown by dashed line also includes the quasielastic like lepton production cross section in the delta dominance model. The experimental points are taken from Armenise et al.~\cite{armenise}.}
\end{figure}
\subsection{Charged current inelastic lepton production}
In this section, we present the numerical results for the total scattering cross sections $\sigma(E)$ vs E, $Q^2$ distributions and lepton angular distributions for the inelastic charged current lepton production process ${\bar{\nu}}_\mu + ^{12}C \rightarrow \mu^+ + \pi^-(\pi^0) + X$ in the $\Delta$ dominance model. The numerical calculations have been done by using Eq.(34) with N-$\Delta$ transition form factors given by Lalakulich et al.\cite{paschos2}. 
\begin{figure}
\includegraphics{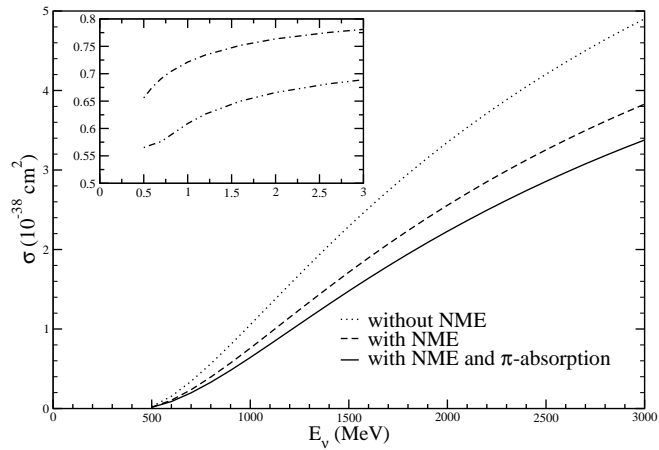}
\caption{The cross section for inelastic charged current lepton production accompanied by a pion induced by
 ${\bar\nu}_\mu$ on $^{12}$C target, with(without) nuclear medium effects(NME) and with nuclear medium and pion absorption effects. In the inset, ratio of the cross section in $^{12}$C as a function of anti-neutrino energy is shown. The dashed-dotted line (dashed double-dotted line) is the result of the ratio when the cross section is calculated including nuclear medium effects with(without) pion absorption effects to the cross section calculated without the nuclear medium effects.}
\end{figure}

In Fig.12, we show the numerical results for the total cross section $\sigma(E)\sim E$ calculated without including nuclear medium effects and including nuclear medium effects with(without) 
pion absorption effects. We find that the
nuclear medium effects lead to a reduction of around 12-15\% for
anti-neutrino energies $\text E_{\bar\nu}$=0.6-3~GeV. When pion absorption
effects are taken into account the total reduction in the cross section is around
$25-30\%$. To quantify our results, in the inset we have presented the ratios of the cross sections calculated including nuclear medium with(without) pion absorption effects to the cross section calculated without including nuclear medium effects. 

In Figs.13 and 14, we have presented respectively the results for the $Q^2$-distribution i.e. $\langle\frac{d\sigma}{dQ^2}\rangle$ vs $Q^2$ and lepton angular distribution i.e. $\langle\frac{d\sigma}{dcos\theta}\rangle$ vs $cos\theta$, averaged over the MiniBooNE spectrum. The results are presented for the differential cross sections calculated without including nuclear medium effects and including nuclear medium effects with(without) pion absorption effects. We find that for $Q^2$ distribution the reduction in the differential cross section with nuclear medium effects is around 15\% in the peak region of $Q^2$. When pion absorption effects are also taken into account the total reduction is around 28-30\%. In the case of angular distribution, we find that at forward angles the reduction in the differential cross section with nuclear medium effects is around 15-20\%. When pion absorption effects are also taken into account the total reduction is around 30\% in the peak region of $Q^2$. In the inset of these figures we have presented the ratios of the differential cross sections calculated including nuclear medium with(without) pion absorption effects to the differential cross section calculated without including nuclear medium effects.\\
\begin{figure}
\includegraphics{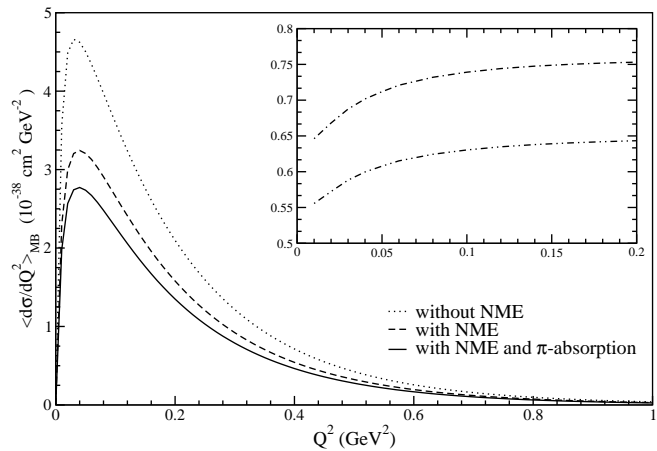}
\caption{$\langle\frac{d\sigma}{dQ^2}\rangle$ vs $Q^2$ for ${\bar\nu}_\mu$ on $^{12}$C target averaged over the MiniBooNE spectrum for the inelastic charged current lepton production process accompanied by a pion, with(without) nuclear medium effects(NME), and with nuclear medium and pion absorption effects. In the inset, ratio of the differential cross section in $^{12}$C as a function of $Q^2$ is shown. The dashed-dotted line (dashed double-dotted line) is the result of the ratio when the cross section is calculated including nuclear medium effects with(without) pion absorption effects to the cross section calculated without the nuclear medium effects.}
\end{figure}
\begin{figure}
\includegraphics{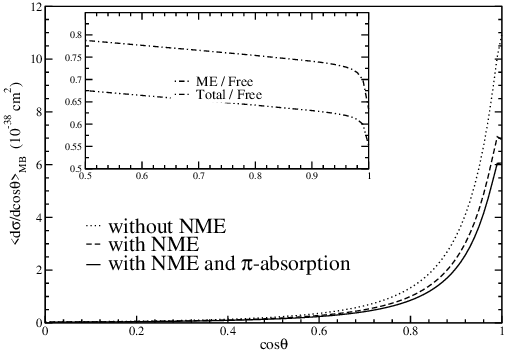}
\caption{$\langle\frac{d\sigma}{dcos\theta}\rangle$ vs $cos\theta$ for ${\bar\nu}_\mu$ on $^{12}$C target averaged over the MiniBooNE spectrum for the inelastic charged current lepton production process accompanied by a pion, with(without) nuclear medium effects(NME), and with nuclear medium and pion absorption effects. In the inset, ratio of the angular distribution in $^{12}$C as a function of lepton angle is shown. The dashed-dotted line (dashed double-dotted line) is the result of the ratio when the cross section is calculated including nuclear medium effects with(without) pion absorption effects to the cross section calculated without the nuclear medium effects.}
\end{figure}

In literature various parameterizations for the N-$\Delta$ transition form factors have been discussed. To study the dependence of the N-$\Delta$ transition form factors on the differential
 cross section $\langle\frac{d\sigma}{dQ^2}\rangle$ vs $Q^2$ we have presented our results in Fig.15 for the various N-$\Delta$ transition form factors
taken from Lalakulich et
al.~\cite{paschos2}, Schreiner and von Hippel~\cite{svh} and
Paschos et al.~\cite{paschos1}. We find that in the peak region of $\langle\frac{d\sigma}{dQ^2}\rangle$ vs $Q^2$ the results obtained by using Lalakulich et al.~\cite{paschos2} and Paschos et
al.~\cite{paschos1} transition form factors are within a few percent($<5\%$) than of the differential cross section obtained by using Schreiner and
von Hippel~\cite{svh} N-$\Delta$ transition form factors. Thus the dependence of the various N-$\Delta$ transition form factors on the differential cross sections is quite small.
\begin{figure}
\includegraphics{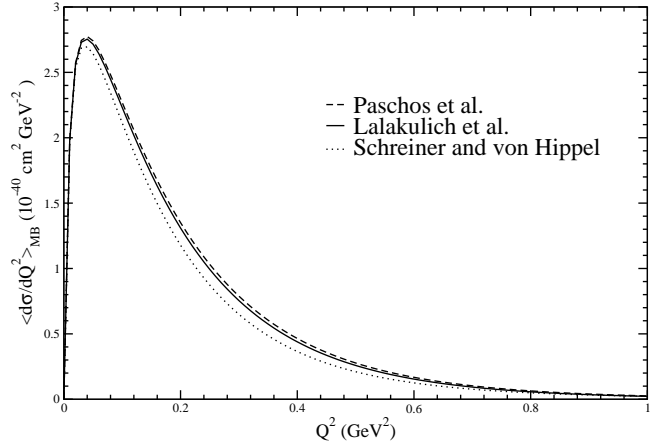}
\caption{$\langle\frac{d\sigma}{dQ^2}\rangle$ vs $Q^2$ calculated with nuclear medium and pion absorption effects for ${\bar\nu}_\mu$ interactions on $^{12}$C target averaged over the MiniBooNE spectrum for the inelastic charged current lepton production process accompanied by a pion using various N-$\Delta$ transition form factors given by Lalakulich et al.\cite{paschos2}, Schreiner and von Hippel~\cite{svh} and Paschos et al.~\cite{paschos1}).}
\end{figure}
\subsection{Charged current incoherent lepton production accompanied by a $\pi^-$}
In this section, we present the numerical results for the total scattering cross sections $\sigma(E)$ vs E, $Q^2$-distributions and lepton angular distributions for the charged current lepton production process accompanied by a $\pi^-$ i.e. ${\bar{\nu}}_\mu + ^{12}C \rightarrow \mu^+ + \pi^- + X$. The numerical calculations have been done using Eq.(49) with N-$\Delta$ transition form factors given by Lalakulich et al.\cite{paschos2}. 
\begin{figure}
\includegraphics{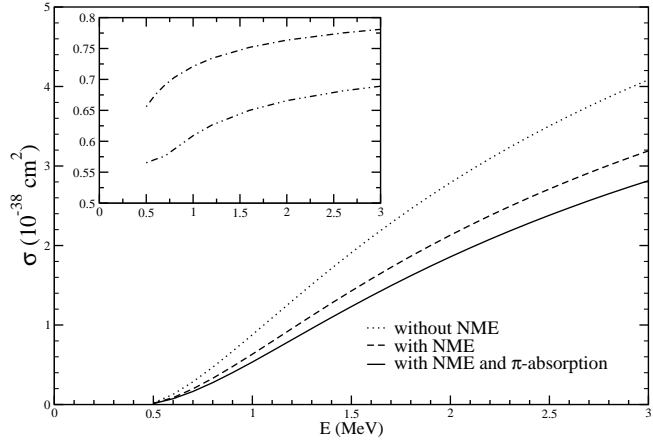}
\caption{Inelastic charged current one $\pi^-$ production cross section induced by
 ${\bar\nu}_\mu$ on $^{12}$C target with and without nuclear medium effects(NME), and with nuclear medium and pion absorption effects. In the inset, ratio of the cross section in $^{12}$C as a function of neutrino energy is shown. The dashed-dotted line (dashed double-dotted line) is the result of the ratio when the cross section is calculated including nuclear medium effects with(without) pion absorption effects to the cross section calculated without the nuclear medium effects.}
\end{figure}

We have shown the numerical results for the total cross section $\sigma(E)$ vs E for the anti-neutrino induced incoherent 1$\pi^-$ production process in Fig.16. The results are presented without including nuclear medium effects and including nuclear medium effects with(without) pion absorption effects. We find that the
nuclear medium effects lead to a reduction of around 12-15\% for
anti-neutrino energies $\text E_{\bar\nu}$=0.6-3~GeV. When pion absorption
effects are taken into account along with nuclear medium
effects the total reduction in the cross section is around
$30-40\%$. In the inset we have presented the ratios of the cross sections calculated including nuclear medium with(without) pion absorption effects to the cross section calculated without including nuclear medium effects. 
\begin{figure}
\includegraphics{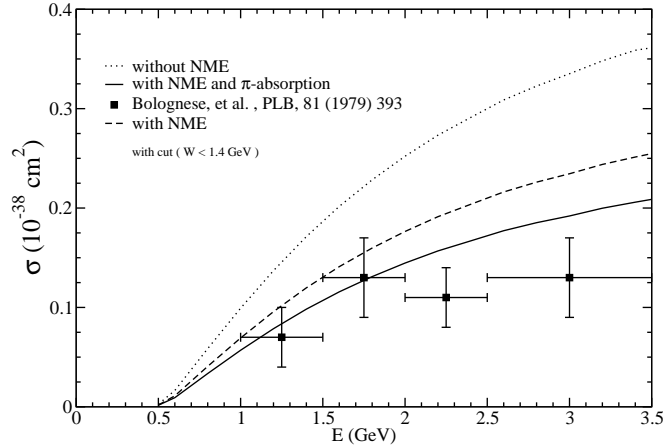}
\caption{Charged current 1$\pi^-$ production cross section per neutron induced by
 ${\bar\nu}_\mu$ on Freon Propane($CF_3Br-C_3H_8$) target, with and without nuclear medium effects(NME), and with nuclear medium and pion absorption effects. The experimental points are taken from Bolognese et al.\cite{bolognese}}.
\end{figure}
\begin{figure}
\includegraphics{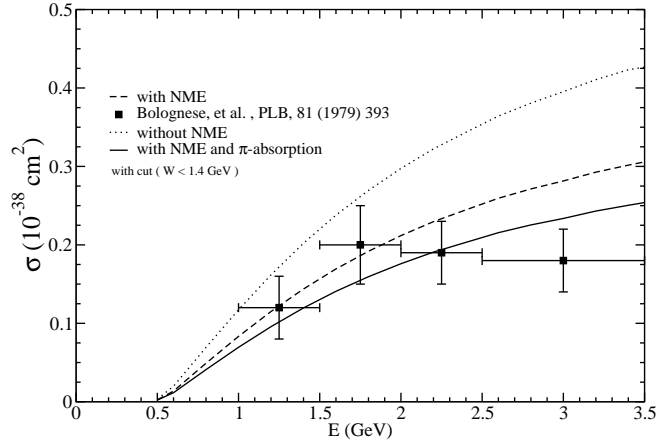}
\caption{Charged current 1$\pi^-$ production cross section per nucleon induced by
 ${\bar\nu}_\mu$ on Freon Propane($CF_3Br-C_3H_8$) target, with and without nuclear medium effects(NME), and with nuclear medium and pion absorption effects. The experimental points are taken Bolognese et al.\cite{bolognese}.}
\end{figure}

In Figs.17 and 18, we present our results for the total cross section obtained for the incoherent 1$\pi^-$ production process on neutron and nucleon targets induced by ${\bar\nu}_\mu$ on Freon-Propane($CF_3Br-C_3H_8$) and compared our results with the experimental results of Bolognese et al.~~\cite{bolognese}. The results of cross section have been presented for total scattering 
cross section $\sigma(E_\nu)$ calculated without including nuclear medium effects and including nuclear medium effects with(without) pion absorption effects. In the case of the cross section calculated for the neutron target in Freon-Propane shown in Fig.17, we find that when cross section is calculated without nuclear medium effects, ${\chi^2}$pdf is 17.7, which reduces to 9.3 when nuclear medium effects are taken into account. Furthermore, when both the nuclear medium and pion absorption effects are included ${\chi^2}$pdf is reduced from 17.7 to 1.3. While in the case of the cross section calculated for the nucleon target in Freon-Propane shown in Fig.18, we find that ${\chi^2}$pdf is 12 which reduces to 5.0 when nuclear medium effects are taken into account. Furthermore, when both the nuclear medium and pion absorption effects are included ${\chi^2}$pdf is reduced from 12 to 0.7. Thus the inclusion of nuclear medium and final state interaction effects leads to a better description of the experimental data.

In Figs.19 and 20, we have presented respectively the results for the $Q^2$-distribution i.e. $\langle\frac{d\sigma}{dQ^2}\rangle$ vs $Q^2$ and lepton angular distribution i.e. $\langle\frac{d\sigma}{dcos\theta}\rangle$ vs $cos\theta$, averaged over the MiniBooNE spectrum for anti-neutrinos. The results are presented for the differential cross sections calculated without including nuclear medium effects and including nuclear medium effects with(without) pion absorption effects. We find that for the $Q^2$ distribution, the reduction in the differential cross section with nuclear medium effects is around 15\% in the peak region of $Q^2$. When pion absorption effects are also taken into account the total reduction is around 35\%. In the case of angular distribution, we find that at forward angles the reduction in the differential cross section when calculated with nuclear medium effects to the differential cross section calculated without nuclear medium effects is around 15-20\%. When pion absorption effects are also taken into account the total reduction is around 40-45\%. In the inset of these figures, we have presented the ratios of the differential cross sections calculated including nuclear medium with(without) pion absorption effects to the differential cross section calculated without including nuclear medium effects.
\begin{figure}
\includegraphics{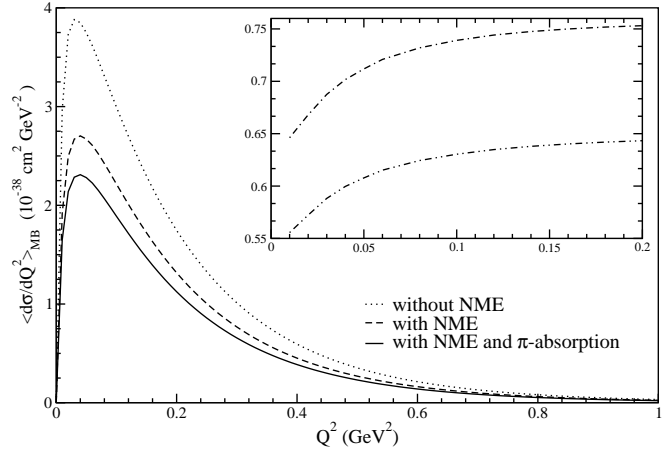}
\caption{$\langle\frac{d\sigma}{dQ^2}\rangle$ vs $Q^2$ for ${\bar\nu}_\mu$ on $^{12}$C target averaged over the MiniBooNE spectrum for the inelastic charged current lepton production process accompanied by a $\pi^-$, with and without nuclear medium effects(NME), and
 with nuclear medium and pion absorption effects. In the inset, the dashed-dotted line (dashed double-dotted line) is the result of the ratio when the cross section is calculated including nuclear medium effects with(without) pion absorption effects to the cross section calculated without the nuclear medium effects.}
\end{figure}
\begin{figure}
\includegraphics{f20.eps}
\caption{$\langle\frac{d\sigma}{dcos\theta}\rangle$ vs $cos\theta$ for ${\bar\nu}_\mu$ on $^{12}$C target averaged over the MiniBooNE spectrum for the inelastic charged current lepton production process accompanied by a $\pi^-$, with and without nuclear medium effects(NME), and with nuclear medium and pion absorption effects. In the inset, the dashed-dotted line (dashed double-dotted line) is the result of the ratio when the cross section is calculated including nuclear medium effects with(without) pion absorption effects to the cross section calculated without the nuclear medium effects.}
\end{figure}
\begin{figure}
\includegraphics{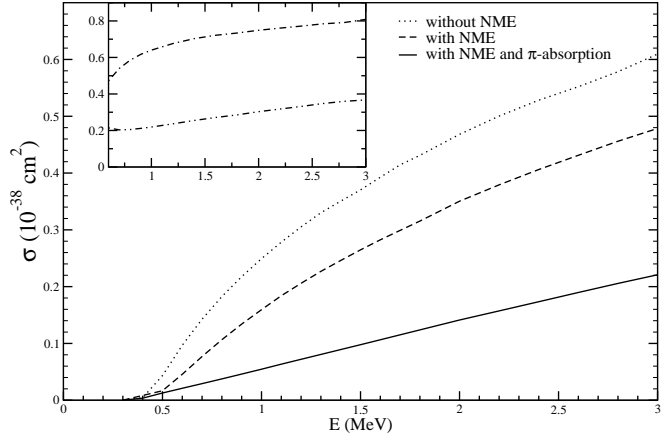}
\caption{Coherent charged current one pion production cross section induced by
 ${\bar\nu}_\mu$ on $^{12}$C target, with and without nuclear medium effects(NME), and with nuclear medium and pion absorption effects. In the inset, the dashed-dotted line (dashed double-dotted line) is the result of the ratio when the cross section is calculated including nuclear medium effects with(without) pion absorption effects to the cross section calculated without the nuclear medium effects.}
\end{figure} 
\subsection{Charged current coherent production of leptons accompanied by a $\pi^-$}
In the case of coherent reactions, the total production of leptons is the same as the leptons accompanied by a $\pi^-$, as in both cases the interacting nucleus remains in the same state(see Eqs.(18) and (19)). In this section, we present the numerical results for the total scattering cross sections $\sigma(E)$ vs E, $Q^2$-distributions and lepton angular distributions for the coherent charged current lepton production process accompanied by a $\pi^-$ i.e. ${\bar{\nu}}_\mu + ^{12}C \rightarrow \mu^+ + \pi^- + ^{12}C$. The numerical calculations have been done using Eq.(47) with N-$\Delta$ transition form factors given by Lalakulich et al.~\cite{paschos2}. 
\begin{figure}
\includegraphics{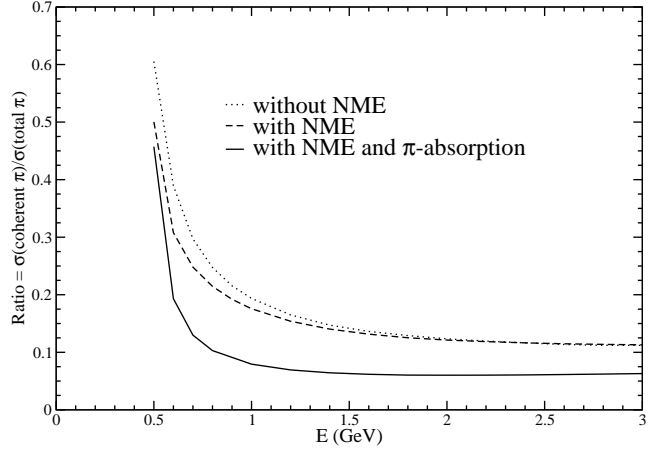}
\caption{Ratio of coherent charged current 1$\pi^-$ production cross section to the total charged current 1$\pi^-$ production cross section(incoherent+coherent) i.e. $\frac{\sigma^{coh}}{\sigma^{total}}$ without nuclear medium effects(NME), with nuclear medium effects and with nuclear medium and pion absorption effects.}
\end{figure}
\begin{figure}
\includegraphics{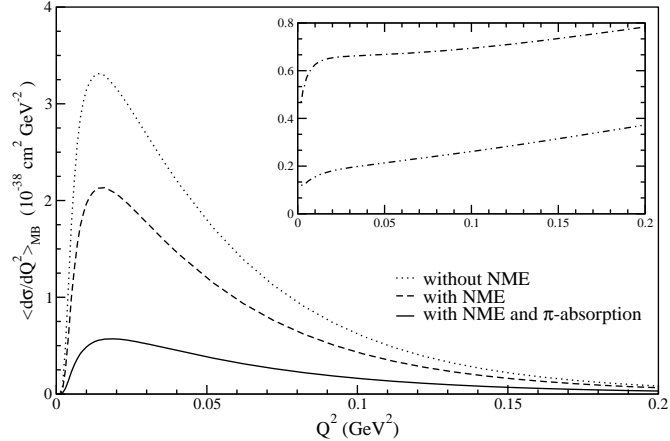}
\caption{$\langle\frac{d\sigma}{dQ^2}\rangle$ vs $Q^2$ for ${\bar\nu}_\mu$ on $^{12}$C target averaged over the MiniBooNE spectrum for the coherent charged current 1$\pi^-$ production process, with and without nuclear medium effects(NME), and with nuclear medium and pion absorption effects. In the inset, the dashed-dotted line (dashed double-dotted line) is the result of the ratio when the cross section is calculated including nuclear medium effects with(without) pion absorption effects to the cross section calculated without the nuclear medium effects.}\end{figure} 

We have shown the numerical results for the total cross section $\sigma(E)$ vs E for the anti-neutrino induced coherent 1$\pi^-$ production process in Fig.21. The results are presented for total scattering
cross section $\sigma(E_\nu)$ calculated without including nuclear medium effects and including nuclear medium effects with(without) pion absorption effects. For the coherent process, the
nuclear medium effects lead to a reduction of around $45\%$ for $\text
E_\nu$=0.7~GeV, $35\%$ for $\text E_\nu$=1~GeV, $25\%$ for $\text E_\nu$=2~GeV. The pion absorption
effects taken into account along with nuclear medium
effects lead to a very large reduction in the total scattering cross
section. The suppression in the total cross section due to nuclear medium
and pion absorption effects in our model is found to be $80\%$ for E$_\nu$ 1 GeV and $70\%$ for E$_\nu$ 2 GeV. Due to large reduction in the total cross
section for the coherent process its contribution to the total
charged current 1$\pi^-$ production is
$<10-12\%$ in the anti-neutrino energy region of 1-2~GeV. This is found to be smaller than the predictions of the NUANCE neutrino generator~\cite{a1}. This contribution is larger than obtained for the neutrino process($<4-5\%$)~\cite{prd1} in our model. This is expected because the coherent cross section remains almost the same for the neutrino and anti-neutrino induced reactions while incoherent production cross section is smaller for the anti-neutrino as compared to the neutrino. We have explicitly shown in Fig.22 the contribution of charged current coherent 1$\pi^-$ production to the total charged current 1$\pi^-$ production(coherent+incoherent) in the $\Delta$ dominance model i.e. $r=\frac{\sigma^{coherent}}{\sigma^{coherent+incoherent}}$. The results are presented for the ratio of the cross sections calculated without including nuclear medium effects and including nuclear medium effects with(without) pion absorption effects. We find that if nuclear medium effects are not taken into account the contribution of the coherent 1$\pi^-$ production to the total 1$\pi^-$ production is 30\% at $E_{\bar\nu}=0.8GeV$, 20\% at $E_{\bar\nu}=1.2GeV$ and 14\% at $E_{\bar\nu}=3.0GeV$. When nuclear medium effects are taken into account this becomes 20\% at $E_{\bar\nu}=0.8GeV$, 16\% at $E_{\bar\nu}=1.2GeV$ and 12\% at $E_{\bar\nu}=3.0GeV$. When pion absorption effects are also taken into account this ratio reduces to 12\% at $E_{\bar\nu}=0.8GeV$, 9\% at $E_{\bar\nu}=1.2GeV$ and around 7-8\% at $E_{\bar\nu}=3.0GeV$.
\begin{figure}
\includegraphics{f24.eps}
\caption{$\langle\frac{d\sigma}{dcos\theta}\rangle$ vs $cos\theta$ for ${\bar\nu}_\mu$ on $^{12}$C target averaged over the MiniBooNE spectrum for the coherent charged current 1$\pi^-$ production process, with and without nuclear medium effects(NME) and with nuclear medium and pion absorption effects. In the inset, the dashed-dotted line (dashed double-dotted line) is the result of the ratio when the cross section is calculated including nuclear medium effects with(without) pion absorption effects to the cross section calculated without the nuclear medium effects.}
\end{figure} 
\begin{figure}
\includegraphics{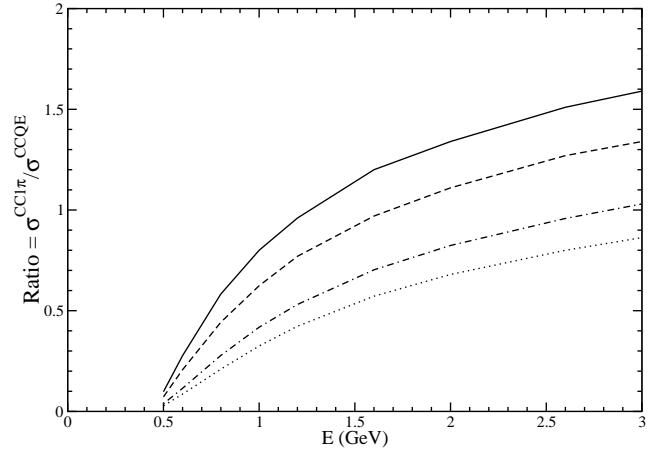}
\caption{Ratio of total 1$\pi^-$ production cross section induced by ${\bar\nu}_\mu$ on $^{12}$C target to the charged current quasielastic lepton production cross section i.e. $\frac{\sigma^{cc1\pi^-}}{\sigma^{ccqe}}$. Result with solid(dashed) line is the ratio of the charged current induced 1$\pi^-$ cross section calculated without nuclear medium effect to the charged current quasielastic cross section in local Fermi gas model with(without) RPA correlations, and the results with dashed-dotted(dotted) line is the ratio of the charged current induced 1$\pi^-$ cross section calculated with nuclear medium and pion absorption effects to the charged current quasielastic cross section in local Fermi gas model with(without) RPA correlations.}
\end{figure} 

In Figs.23 and 24, we have presented respectively the results for the $Q^2$-distribution i.e. $\langle\frac{d\sigma}{dQ^2}\rangle$ vs $Q^2$ and lepton angular distribution i.e. $\langle\frac{d\sigma}{dcos\theta}\rangle$ vs $cos\theta$, averaged over the MiniBooNE spectrum. The results are presented for the differential cross sections calculated without including nuclear medium effects and including nuclear medium effects with(without) pion absorption effects. We find that in the peak region of $\langle\frac{d\sigma}{dQ^2}\rangle$ the reduction in the differential cross section when calculated with nuclear medium effects is 35\% compared to the differential cross section calculated without nuclear medium effects. When pion absorption effects are also taken into account the total reduction is around 85\%. In the case of angular distribution, we find that at forward angles the reduction in the differential cross section is 30\% compared to the differential cross section calculated without nuclear medium effects. When pion absorption effects are also taken into account the total reduction is around 75\%. To explicitly visualise these nuclear medium effects, we have given the ratio of the differential cross section calculated with nuclear medium and with(without) pion absorption effects to the differential cross section calculated without nuclear medium effects in the inset of these figures. 

Recently, MiniBooNE collaboration~\cite{boone1} have reported the ratio $R^\prime=\frac{\sigma(CC1\pi^+)}{\sigma(CCQE)}$ for neutrino induced processes. We have studied this ratio in our earlier paper~\cite{prd1} and found that the theoretical predictions for the cross sections in our model are in reasonable agreement with the experimental results for the ratio reported by them. In Fig.25, we have studied the ratio $R=\frac{\sigma(CC1\pi^-)}{\sigma(CCQE)}$ for anti-neutrino induced processes. We have presented our results for this ratio by using the charged current quasielastic scattering cross sections $\sigma(CCQE)$ obtained in the local Fermi gas model including nuclear medium effects with(without) RPA correlations. For the incoherent and coherent 1$\pi^-$ production process, the cross sections $\sigma(CC1\pi^-)$ are calculated without nuclear medium effects and with nuclear medium and pion absorption effects. We find that when $\sigma(CC1\pi^-)$ is calculated without nuclear medium effects and $\sigma(CCQE)$ is calculated in the local Fermi gas model without RPA correlations, the ratio R is 0.44 at $E_{\bar\nu}=0.8GeV$, 0.77 at $E_{\bar\nu}=1.2GeV$ and 1.34 at $E_{\bar\nu}=3.0GeV$. When RPA correlations are taken into account in calculating $\sigma(CCQE)$ this ratio becomes 0.58 at $E_{\bar\nu}=0.8GeV$, 0.96 at $E_{\bar\nu}=1.2GeV$ and 1.6 at $E_{\bar\nu}=3.0GeV$. However, when $\sigma(CC1\pi^-)$ is calculated with nuclear medium and pion absorption effects and $\sigma(CCQE)$ is calculated in the local Fermi gas model without RPA correlations, this ratio is 0.21 at $E_{\bar\nu}=0.8GeV$, 0.42 at $E_{\bar\nu}=1.2GeV$ and 0.86 at $E_{\bar\nu}=3.0GeV$. Our final results for this ratio when $\sigma(CC1\pi^-)$ is calculated with nuclear medium and pion absorption effects and $\sigma(CCQE)$ is calculated in the local Fermi gas model with(without) RPA correlations, is 0.28 at $E_{\bar\nu}=0.8GeV$, 0.53 at $E_{\bar\nu}=1.2GeV$ and 1.03 at $E_{\bar\nu}=3.0GeV$. Thus, we find that this ratio strongly depends on nuclear medium effects both for the cross sections calculated in the case of the charged current quasielastic lepton production cross section and the inelastic lepton production cross section.
\section{Summary and Conclusions}
The anti-neutrino induced charged current quasielastic and inelastic reactions from nuclei have been studied. These studies are important in the low and intermediate energy region to quantify the role of nuclear medium effects and are useful in the analysis of anti-neutrino experiments being done on nuclear targets like $^{12}C$ at MiniBooNE. We have presently studied the effects of nuclear medium on the total scattering cross section for the charged current anti-neutrino induced quasielastic and inelastic lepton production, incoherent and coherent 1$\pi^-$ production, and the $Q^2$-distribution and the angular distribution of leptons. The calculations have been done in the local Fermi gas model. The effect of Pauli blocking, Fermi motion of the nucleons, Q-value of the reaction and the Coulomb distortion effects on the outgoing lepton have been included. The renormalization of weak transition strengths on the weak couplings i.e. RPA correlations in the nuclear medium has also been taken into account for the quasielastic lepton production process. For the inelastic reactions, the calculations have been done in the $\Delta$ dominance model. The modification of the mass and width of the $\Delta$ resonance in the nuclear medium has been taken into account. Moreover, the final state interaction of pions with the residual nucleus has also been taken into account. The results for the total scattering cross section have been compared with the experimental results of some earlier experiments in Freon and Freon-Propane in the case of quasielastic process and in Freon-Propane in the case of incoherent 1$\pi^-$ production process.

From this study we conclude that:

1. In the case of charged current quasielastic lepton production, the role of nuclear medium effects like Pauli blocking, Fermi motion is to reduce the cross section. When RPA correlations are taken into account there is further reduction in the cross section. We find that the total reduction in the cross section is around 50\% at $E_{{\bar\nu}_\mu}$=0.5GeV, which decreases with the increase in anti-neutrino energy and becomes 25\% at $E_{{\bar\nu}_\mu}$=3.0GeV as compared to the cross sections calculated without nuclear medium effects. Besides the genuine quasielastic lepton events there is also contribution from the inelastic process through $\Delta$ excitation where a $\Delta$ is absorbed in the nuclear medium via the processes like $\Delta$N$\rightarrow$ NN and $\Delta$NN$\rightarrow$ NNN. The contribution of such quasielastic like events is $\sim$8\% at the anti-neutrino energies of 1GeV.

2. The results of total scattering cross section $\sigma(E) \sim E$ for the quasielastic lepton production (including the quasielastic like events from the $\Delta$ excitation) have been compared with the experimental results of Bonetti et al.~\cite{bon} and Brunner et al.~\cite{bru} in Freon($CF_3Br$) and in Freon-Propane by Armenise et al.\cite{armenise}. We find that inclusion of nuclear medium effects lead to good agreement with the experimental data.

3. In the case of charged current inelastic lepton production the inclusion of nuclear medium effects like modification in the mass and width of the $\Delta$ lead to a reduction in the total cross section which is around 12-15\% for
anti-neutrino energies of $E_{{\bar\nu}_\mu}$=0.6-3GeV. When pion absorption effects are also taken into account the total reduction in the cross section is around 25-30\% as compared to the cross sections calculated without the nuclear medium and final state interaction effects. 

4. In the case of charged current inelastic 1$\pi^-$ production the inclusion of the nuclear medium effects are lead to the reduction in the total cross section which is around 12-15\% for
anti-neutrino energies of $E_{{\bar\nu}_\mu}$=0.6-3GeV. When pion absorption effects are also taken into account the total reduction in the cross section is around 30-40\% as compared to the cross sections calculated without the nuclear medium and final state interaction effects. For the coherent 1$\pi^-$ production, the reduction in the total cross section is very large due to nuclear medium and pion absorption effects. Due to large reduction in the total cross section for the coherent process its contribution to the total charged current 1$\pi^-$ production is $<10-12\%$ in the anti-neutrino energy region of 1-2~GeV which is smaller than found in earlier studies.

5. The results of total scattering cross section $\sigma(E) \sim E$ for the incoherent 1$\pi^-$ production have been compared with the experimental results of Bolognese et al.~\cite{bolognese} in Freon-Propane. We find that with nuclear mediums and pion absorption effects our present results are in better agreement with the experimental data.

6. The results for $\langle\frac{d\sigma}{dQ^2}\rangle$ vs $Q^2$ and $\langle\frac{d\sigma}{dcos\theta}\rangle$ vs $cos\theta$ in $^{12}${C} averaged over the MiniBooNE flux for the anti-neutrino reactions have been presented for the quasielastic and inelastic processes. The results have also been presented for $\langle\frac{d\sigma}{dQ^2}\rangle$ vs $Q^2$ using different parameterizations of N-$\Delta$ transition form factors. In the case of anti-neutrino induced charged current inelastic reactions, the dependence of the differential cross sections on the various parameterization of N-$\Delta$ transition form factors are found to be small. 
\section{Acknowledgment}
We would like to thank M. J. Vicente Vacas for providing us the pion absorption probabilities. The work is financially supported by the Department of Science and
Technology, Government of India under the grant DST Project
No. SP/S2K-07/2000.

\end{document}